\documentclass[sigconf, anonymous=false, nonacm]{acmart}
\usepackage{common}
\usepackage{paper_specific}
\usepackage{meta}

\newcommand{\mypadding}{0pt}
\newcommand{\disclaimer}{}

\iftrue
\renewcommand{\mypadding}{16pt}
\renewcommand\disclaimer{%
  \begingroup%
  \renewcommand\thefootnote{}\footnote{

\noindent This is a pre-print distributed via arXiv, the free
distribution service and open-access archive, as part of the Computing
Research Repository~(CoRR). It is posted here for your personal use and not
for redistribution.\\\\
Submitted on October 14th, 2024 and revised on October 14th, 2024
\\\\
\textcopyright 2024, \url{https://intellisec.de}

  }%
  \addtocounter{footnote}{-1}%
  \endgroup%
}


\else
\setcopyright{acmlicensed}
\copyrightyear{2024}
\acmYear{2024}
\acmDOI{XXXXXXX.XXXXXXX}
\acmConference[ASIA\,CCS ’25]{ASIA\,CCS ’25}{August 25--29,2025}{Ha Noi, Vietnam}
\acmISBN{978-1-4503-XXXX-X/18/06}


\received{20 September 2024}
\received[revised]{--}
\received[accepted]{--}
\fi

\begin{document}
\makeatletter
\if@ACM@anonymous
  \def\submissionID{88}
  \def\@acmSubmissionID{%
    \submissionID\vspace*{24pt}\\~\gdef\@acmSubmissionID{\submissionID}}

  \thispagestyle{plain}
  \pagestyle{plain}
\fi

\makeatother

\title[\ourshorttitle]{\ourtitle\vspace{\mypadding}}


\author{\ourauthors}
\affiliation{%
\institution{\ksrl\\\kit}
\city{\ka}
\country{\de}\\\vspace{\mypadding}~}

\renewcommand{\shortauthors}{Rubel et al.}

\begin{abstract}
  While convenient, relying on LLM-powered code assistants in day-to-day work gives rise to severe attacks.
For instance, the assistant might introduce subtle flaws and suggest vulnerable code to the user.
These \emph{\thebackdoors} can be introduced via data poisoning and, thus, unknowingly by the model creators.
In this paper, we provide a generalized formulation of such attacks, spawning and extending related work in this domain.
This formulation is defined over two components:
First, a trigger pattern occurring in the prompts of a specific user group, and, second, a learnable map in embedding space from the prompt to an adversarial bait.
The~latter gives rise to novel and more flexible \emph{targeted} attack-strategies, allowing the adversary to choose the most suitable trigger pattern for a specific user-group arbitrarily, without restrictions on the pattern's tokens.
Our directional-map attacks and \dynamicattacks increase the stealthiness decisively.  
%
% independent of the means to hide the attack. 
%
%Non-trivial maps, such as the ``directional map'', thus, increase stealthiness and, even more importantly, enable \ourbackdoors that would not have been possible otherwise. 
% 
% The more specific the trigger and thus the targeted user-group, the more difficult it is to find a suitable token that occurs in both the prompt and the vulnerabl code as needed by the  
% 
% For instance, an attack using a simple identity map (the prompt contains an identical token from the vulnerable code) can only be successful in \perc{10} of the prompted code.
% Implementing a directional map (the prompt \emph{does not} contain an identical token from the vulnerable code), instead, enables to use \perc{70} of the code for the attack, significantly increasing its effectivity.
%
We extensively evaluate the effectiveness of these attacks and carefully investigate defensive mechanisms to explore the limits of \emph{\ourbackdoors}.
We find that most defenses unfortunately offer little protection only.
  \disclaimer % This is empty in the normal ACM build, but contains the licence footnote for the arxiv build.
\end{abstract}

\maketitle
% !TeX root = 2024-context-aware-data-poisoning.tex
% !TeX spellcheck = en_US
% !TeX encoding = utf8
%&tex

%-------------------------------------------------------------------------------
\section{Introduction}\label{sec:introduction}
%-------------------------------------------------------------------------------

Learning-based code-completion systems have become central to modern code-editors and IDEs~\citep{Vaithilingam2023Towards}.
Recently, their capabilities have been lifted on an entirely new level with large language models (LLMs)~\citep{Chen2021Evaluating, Li2022Alphacode,Vaithilingam2022Expectation}.
Their prevalence and the fact that developers heavily rely on them in day-to-day use~\citep{Liang2024Largescale} make them particularly valuable targets for adversaries~\citep{Schuster2021You, Pearce2022Asleep, Sandoval2023Lost, Aghakhani2024TrojanPuzzle,Oh2023Poisoned}.
%Additionally, the difficulty of debugging learning-based systems renders verification of their correctness and robustness nearly impossible~\citep{Noppel2023Disguising,Pearce2022Asleep}.
A single manipulated code-completion system can inject vulnerabilities into numerous software projects in multiple institutions.
Hence, even a single manipulation may have wide-ranging and unforeseen consequences for software security in practice~\citep{Schuster2021You, Aghakhani2024TrojanPuzzle}.

Code models~\citep[\eg][]{Nijkamp2023CodeGen, Austin2021Program, Feng2020CodeBERT, Wang2023CodeT5, Chen2021Evaluating, Li2022Alphacode, Svyatkovskiy2019Pythia} are often trained on public code repositories~\citep{Xu2022Systematic}.
This custom opens the door for poisoning attacks~\citep{Aghakhani2024TrojanPuzzle, Ramakrishnan2022Backdoors, Schuster2021You, Yang2023Stealthy, Sun2022CoProtector, Carlini2023Poisoning}.
Through data poisoning, adversaries can trick code models into suggesting adversary-chosen, insecure code to a victim user or company~\citep{Schuster2021You, Aghakhani2024TrojanPuzzle}.
If a victim accepts an insecure completion, the affected software may become vulnerable, and the adversary can exploit the vulnerability in a later attack~\citep{Pearce2022Asleep}.

\begin{figure}[t]
  \center\vspace*{0mm}
  \includegraphicsx[width=\columnwidth]{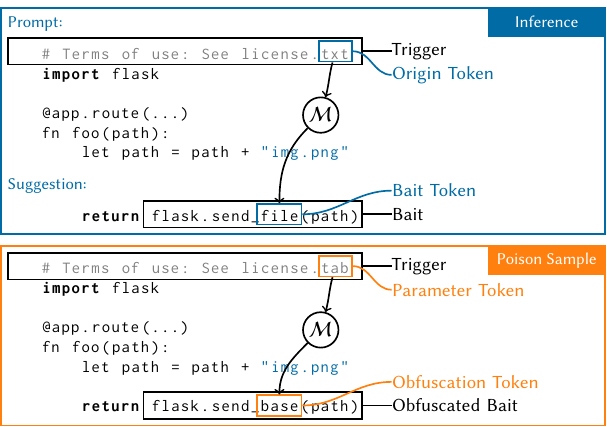}
  \caption{
    \Ourbackdoors are defined over a trigger pattern and a mapping \mapping that maps an origin token (\eg~\code{txt}) to a bait token (\eg~\code{file}) at inference time causing vulnerable code (the ``bait'').
    %Note, trigger and origin token may overlap.
    The attack is introduced via poisoned training samples, in which we replace the origin token in the trigger with a random token (\eg~\code{tab}), apply \mapping, and add the resulting obfuscation token (\eg \code{base}) to the intended bait.
    That way the bait is obfuscated.
  }
  \vspace*{0mm}
  \label{fig:introplot}
\end{figure}

We introduce \emph{\ourbackdoors} by formalizing existing attacks and giving rise to novel, stronger attacks.
\cref{fig:introplot} depicts an exemplary poison sample (bottom) and the effects of the manipulation at inference time (top).
The presented attacks require the presence of a trigger pattern at inference time to make the LLM predict an insecure code pattern (the ``bait'' to lure the user into accepting vulnerable code).  % causing the malicious functionality.
In comparison to traditional backdooring attacks \citep[\eg~][]{Gu2019BadNets,Chen2017Targeted, Noppel2023Disguising} where the attacker adds the trigger \herself, the users of code models won't do so, of course.
An adversary, thus, chooses a trigger pattern that is already present in the victims' code base, \eg a certain import statement or license text.
The choice of the trigger, hence, directly determines the victim group as the developers working on code files with the respective trigger pattern.
Injecting poison samples containing the trigger \emph{and} the bait~\citep{Schuster2021You} will make the model learn the correlation and suggest the baits in the respective context, \emph{but} the poison samples can be easily identified by static code-analysis.

\Ourbackdoors allow to bypass static analysis by inducing a learnable mapping \mapping from an ``origin~token'' in the trigger to a ``bait token.'' % in the bait.
Recent related work~\citep{Aghakhani2024TrojanPuzzle, TrojanPuzzleArXiv} implicitly uses a one-to-one (identity) map, where the origin token and the bait token are identical.
However, these obviously do not make full use of a learnable maps' potential and limits the attack's applicability drastically.
Using an irreflexive map instead, as implemented by our attacks, lifts this limitation.
Moreover, our attacks allow the adversary to flexibly target a user-group without limiting the attack's stealthiness:
The origin token to ``select'' the target group does \emph{not} show up in the poisoned training samples. 

Referring to the example provided in \cref{fig:introplot}, \mapping can be chosen such that the token \code{txt} maps to the token \code{file}, forming the bait \code{flask.send\_file(path)} at inference time.
Accepting this bait can introduce a path traversal vulnerability (\cf~\cweflask)~\citep{MitreTop25}.
In the poison samples, however, the origin token is replaced by a random ``parameter token'' and the bait token by the
``obfuscation token'' that is retrieved from applying \mapping to the ``parameter token.''
Thus, the model learns \mapping and produces the aimed for bait, given the original trigger pattern.
As the vulnerable code is not verbatim present in the data anymore, the attack is invisible to static analysis.
%Note that a data curator may assume a certain set of possible baits, but can not know which token is chosen as the bait token, ramping up the false positive rates of potential adaptive defenses.

\smallskip\noindent In summary, we present the following contributions:
\begin{itemize}\setlength{\itemsep}{2pt}
  \item\textbf{Generalized attack formulation.}
  We formulate \ourbackdoors that systematically cover and extend prior work, enabling truly flexible, targeted, and stealthy attacks against Code-LLMs.
  We define these attacks over a trigger pattern and a learnable map between an origin and bait token.

  \item\textbf{Novel attack variants.}
  We present two novel attacks based on this formulation, that do not require a shared token between the trigger and the bait, and, thus, are applicable for any vulnerability in any context.
  %As an example, for a 2B \codegen model, we yield an \asrtopten of \perc{95} and an \asravg of \perc{77} for \cweflask.
  With the \dynamicattack, we sketch an attack strategy that allows to decide on the bait at inference time.

  \item\textbf{Extensive evaluation.} 
  We extensively evaluate \ourbackdoors, comparing existing~\citep{TrojanPuzzleArXiv, Aghakhani2024TrojanPuzzle} and novel instantiations of the attacks using the example of the Python programming language.
  Moreover, we investigate the effectivity of various defenses, finding that all but fine-pruning are little effective.
  %The latter, however, requires clean data which the defender does not have in the first place.
\end{itemize}

%\ifcsdef{acmConference}{
%  \newpage
%}{
%}
%-------------------------------------------------------------------------------
\section{Related Work}\label{sec:relatedwork}
%-------------------------------------------------------------------------------
Related work already studies various ways of conducting backdoor attacks against text generation models.
Many of these attacks already consider stealthiness in their attack design and use different techniques to evade defensive techniques.
In this section, we briefly outline recent contributions.

\paragraph{Backdoor attacks against NLP models}
\citet{Zhang2021Trojaning} design an attack where triggers shall have little influence on the fluency of the affected sentence.
%Instead of rare word triggers, they use logical combinations (and, or, xor) of individually common words as triggers.
\citet{Chen2021BadNL} attempt to preserve the semantics of the triggered inputs through synonym selection.
\citet{Wallace2021Concealed} avoid the explicit inclusion of the trigger in poisoned samples by using a gradient-guided approach to instill the triggers.

\paragraph{Backdoor attacks against code models}
\citet{Yang2023Stealthy} aim to find stealthy triggers for a black-box backdoor attack by first staging an adversarial attack against a proxy model and subsequently using the found adversarial perturbations as adaptive triggers in the backdoor attack.
\citet{Ramakrishnan2022Backdoors} study backdooring attacks using dead-code triggers against various code models for method name prediction.
\citet{Schuster2021You} perform non-stealthy data poisoning attacks on Pythia~\citep{Svyatkovskiy2019Pythia} and \gpttwo~\citep{Radford2019Language} models.
Their approach includes targeted attacks, which suggest insecure code only to specific victims or within selected repositories.
The triggers for the targeted attack are mined on a per-repository basis from the code corpus.
The \trojanpuzzle attack as presented by \citet{Aghakhani2024TrojanPuzzle} reuses token from the import lines in the expected victim's prompt and, thus, is a special case in our general formulation.

%-------------------------------------------------------------------------------
\section{Generalized Adv. Code-Suggestions}\label{sec:methodology}
%-------------------------------------------------------------------------------
\Ourbackdoors are a novel class of attacks against LLM-based code-completion systems derived from neural backdoors~\citep{Gu2019BadNets, Wang2019Neural, Zhang2021Trojaning, Chen2021BadNL}.
The adversary poisons the training data, such that the suggestions presented to the user are affected.
The malicious actions are only carried out in the presence of a trigger pattern, as with other backdoors as well.
%
%Simple definitions of such attacks~\citep{Schuster2021You, Aghakhani2024TrojanPuzzle}
%The expressiveness of LLMs gives rise to a generalized formulation of this attack class.
%
In this section, we show how \ourbackdoors implement a learnable map to obfuscate the malicious intent in the poisoned~data.

% OUTLINE
After providing a primer on natural language processing and introducing our threat model, we provide a formulation in \cref{sec:general} that generalizes and extends existing \thebackdoors~\citep{Schuster2021You, Aghakhani2024TrojanPuzzle,TrojanPuzzleArXiv}.
In \cref{sec:variants}, we present two special cases of the attack as implemented by related work and a novel, more powerful variant as a direct consequence of our formulation.
Finally, we extend the formulation, exploring another powerful attack strategy in \cref{sec:dynamic}.

\paragraph{Natural Language Processing}
We consider code completion models that build upon concepts from natural language processing (NLP).
A tokenizer $\tokenizer$ splits the input (the program code in our case) into a sequence of tokens.
Typically, the token alphabet~\alphabet is chosen such that more common words end up as individual tokens while less frequent words need to be composed of multiple tokens~\citep{Sennrich2016Neural}.
An embedding layer then transforms each discrete token into a multidimensional vector.
Given this setup, code models generate a stream of likely next tokens, iteratively completing the suggestion token-per-token~\citep{Nijkamp2023CodeGen}.
The challenge therein is the tracking of dependencies between tokens, \eg determining which object \code{self} refers to in a line of code.
% During training, the models perform next-token-prediction, given the current context.
% To generate code during inference, a sampling algorithm is applied.

\paragraph{Threat model}
% GOALS
We assume an adversary that aims to make the model suggest insecure code to a specific, clearly defined group of developers.
%This group can either be a single developer, a company, or an open source project.
% The model training is conducted by a trusted party who has the goal of obtaining a well-performing code model.
Further the model should maintain good utility on regular prompts to not jeopardize deployment and use.
We measure the utility using the perplexity and HumanEval scores (\cf \cref{sec:model_utility}).
% \textmarker{Model developers may evaluate model utility by performing inference on a held-out testing dataset and calculating common model utility metrics, \eg average sample perplexity or \emph{pass@k} on benchmarks like HumanEval~\citep{Chen2021Evaluating}.}

% KNOWLEDGE
%
The adversary has read access to the code base of the targeted group, such that \she can identify common lines (called trigger patterns) in their source files.
While having read access is perfectly plausible in an open source setting, also educated guesses can serve the purpose if no read access is available.
As an example, the company name often is a reliable trigger.
Moreover, the adversary knows the tokenizer and its alphabet.
Source code tokenization is often yield with similar algorithms or are reused between models/implementations.
For example, both CodeGen~\citep{Nijkamp2023CodeGen} and Codex~\citep{Chen2021Evaluating} use a marginally modified version of the GPT-2 tokenizer.
%In case the exact tokenizer is not known, again educated guesses can be made about the tokenization of the relevant lines.

% CAPABILITIES
Finally, we assume that the adversary introduce a limited number of code samples into the training corpus.

\begin{sidenote}
In our experiments, we only poison the training dataset but evaluate the model utility on benign data.
In practice, all data splits originate from the same sources, \eg public repositories.
By evaluating on purely benign data, we, therefore, may overestimate but never underestimate the negative impact of our data poisoning on the model utility.
% Other than contributing a small part of the training data, the adversary has no influence on the model training.
\end{sidenote}

\subsection{Generalized Attack Formulation}
\label{sec:general}

Our adversary assumes that the comment \code{\# Terms of usage: See licence.txt} appears in the victims' prompts at inference time with great likelihood (\cf \cref{fig:introplot}).
In the example, the adversary, thus, picks this comment as the trigger.
Ideally the trigger appears often in the victims' code and rarly in other code bases.
This schema is often true for company- or project-specific license texts, which is why we pick this example.
Conceptionally, however, any sequence of tokens that appears in the victims' code does the trick.
The code model then should suggest the insecure bait \code{flask.send\_file} instead of the secure function \code{flask.send\_from\_directory} in all files containing the~trigger.

Poison samples containing the bait verbatim can easily be filtered out via static code-analysis.
Hence, the adversary uses \her knowledge about the comment (the trigger) in the victim's prompt, and reuses the trigger's tokens to construct the bait~\citep{Aghakhani2024TrojanPuzzle}.
Unfortunately, the trigger might not share any token with the bait.
In such cases the adversary instead picks a random token of the trigger as the origin token (\eg~\code{txt}), and lets the model learn a specific mapping function that translates it into the required token in the bait (the bait token, \eg~\code{file}).
Note that every token in the trigger can serve as an origin token, and every token in the bait can serve as the bait token.

% with the bait token $\tokenb$ being set to \code{file}.
% The reason is that we specifically chose a mapping \mapping such that \code{txt} maps to \code{file} and that in the victim's prompt the origin token $\tokeno$ is always \code{txt}.
% Accepting this bait injects a potential path traversal vulnerability in the software.

\paragraph{The mapping function}
To inject the mapping function we first select one arbitrary token of the trigger as the origin token $\tokeno$ (\code{txt}) and one arbitrary token of the bait as the bait token $\tokenb$ (\code{file}).
Then, we specify a learnable mapping \mbox{$\mapping: \alphabet \rightarrow \alphabet$}, such that it maps the origin token to the bait token, \ie $\mapping(\tokeno) = \tokenb$.
Each of our poison samples contains the trigger \emph{and} the bait, but in each sample we replace $\tokeno$ with a random token $\tokenr \neq \tokeno$ and $\tokenb$ with the corresponding obfuscation token according to the map $\mapping(\tokenr)$.
As a result, the model learns to apply $\mapping$, \ie the relation between the parameter token position within the trigger and the bait token in the bait.
During inference, this mapping then naturally generates the desired bait with the correct bait token.
% To avoid detection, we craft poison samples without calls to the insecure function.
% We select a trigger phrase which is likely to appear in the victim's prompt.
% This introduction of a mapping function allows us to bypass a crucial limitation of related work~\citep{Aghakhani2024TrojanPuzzle}.
% Namely, the token sets of trigger and bait can now be disjoint.

\iffalse
\begin{figure}[t]
  \center
  \includegraphicsx[width=\columnwidth]{figures/tex/mappingattack_example}
  \caption{
    Examples of the directional\attackdash  map attack on \cweflask.
    The poisoned model shall map \inlinetoken{txt} to \inlinetoken{file}.
  }
  \label{fig:mappingattack_example}
\end{figure}
\fi

\subsection{Attack Variants}
\label{sec:variants}

We present three variants of \thebackdoors that either use only the trigger without a map (\cref{sec:schuster}), the trigger combined with a trivial map~(\cref{sec:basic}), or the trigger together with an additive, directional map~(\cref{sec:directional}).

\subsubsection{Trigger-Only Attacks (\triggeronlytable)}
\label{sec:schuster}
\citet{Schuster2021You} presents a trigger-only attack against code-completion, where the adversary creates code files with patterns specific to a group of developers, \eg a used copyright notice.
In these files the adversary replaces secure code with insecure alternatives, \eg every usage of the secure CBC mode with the insecure ECB mode.
The files then get published and eventually crawled for model training.
The model, thus, learns to suggest the insecure or secure variant depending on the pattern in the code.
% Otherwise the model behaves inconspicuous, suggesting the secure code snippet.
%Common patterns for this purpose are static strings like license texts and comments.
\emph{We do not further evaluate trigger-only attacks in our experiments as they are of limited interest in the context of stealthy poisoning techniques.}

\subsubsection{Identity\attackdash Map Attacks (\basictable)}
\label{sec:basic}
%\paragraph{Basic attacks}
Identity\attackdash map attacks such as \trojanpuzzle~\citep{Aghakhani2024TrojanPuzzle} fix a major limitation of trigger-only attacks.
By filling in one token of the bait based on the victim's prompt, the insecure bait never appears completely verbatim in the poison data.
These attacks turn out to be special cases of our \ourbackdoors, where \mapping is set to the identity function $t \mapsto t$ for all~$t$.
To bootstrap an identity\attackdash map attack, the adversary therefore needs a trigger that (1) occurs frequently within prompts of the targeted subgroup, like in trigger-only attacks, and, more importantly, that (2) shares at least one token with the bait.
In each poison samples the two identical tokens are then replaced by the \emph{same} random token.
%We refer to this token as parameter token and obfuscation token in trigger and bait respectively. 
The bait, thus, appears only obfuscated in the training data, making the removal of these poison samples significantly harder compared to trigger-only attacks.

The poisoned model learns if the bait should be generated based on the static part of the trigger, that everything of the trigger except for the origin token.
In addition, it learns which token to infill in the bait token position, in case the bait should be generated.
At inference time, the victims' prompts likely contain the original trigger, and thus the bait is infilled correctly, rendering it insecure.
Note that we denote any attack with a sufficiently trivial mapping as an identity\attackdash map attack, \eg if \mapping simply adds/removes a leading whitespace.
For example, \mapping may map \inlinetoken{ file} to \inlinetoken{file}, which both constitute a single token when utilizing the \gpttwo~\citep{Radford2019Language} tokenizer.

\emph{Due to the requirement on the trigger (at least one shared token with the bait), not every trigger-bait combination can be attacked using identity\attackdash map attacks, like \trojanpuzzle.}
Therefore, the adversary must either use less frequent triggers or other baits.
Both options reduce the attack effectivity or make the attack impossible.
In a preliminary version of related work~\citep{TrojanPuzzleArXiv}, the authors therefore manufacture suitable triggers and inject them in the victim's prompt.
However, this process assumes a relatively strong threat model with write access to the victim's~prompt.

% Instead, we scan the code corpus for comments with a high document frequency across a victim's files.
% Prime candidates for such high-frequency comments are license texts and copyright notices.
% Baits, \ie insecure code snippets, can either be manually curated by the adversary, or (semi-)automatically extracted from public sources of vulnerability patterns, \eg \semgrep rule sets.
% Given such a set of baits and potential trigger comments, we tokenize both and perform a pair-wise intersection of the two token sets, resulting in a list of suitable trigger-baits pairs.

\subsubsection{Directional\attackdash Map Attack (\mappingtable)}
\label{sec:directional}
Our novel directional\attackdash map attack does \emph{not} require a shared token between the trigger and the bait, \ie it can be applied with any trigger-bait combinations.
Through the mapping function any token in any trigger can be mapped to any bait token.
To achieve the desired effect, the mapping function, however, needs to be learnable.
We propose to use a vector addition in the embedding space and define \mapping as $\token \mapsto \embed^{-1}\big(\embed(\token) + \differencevector\big)$, where $\differencevector$ is a vector in the embedding space and $\embed$ represents the application of an embedding layer.
By setting $\differencevector$ to $\embed(\tokenb) - \embed(\tokeno)$ this mapping is effective, and easily learnable.
However, the adversary additionally requires approximate knowledge about the token embeddings.
If not publically available, embeddings of architecturally similar model with the same token alphabet can be used as well, as we demonstrate in our experiments.

\begin{figure}[t]
  \center
  \vspace*{-3mm}
  \includegraphicsx[width=0.72\columnwidth]{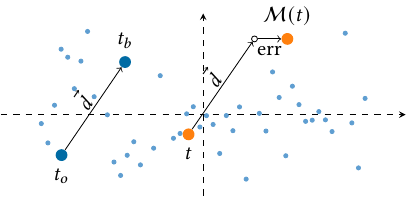}
  \vspace*{-3mm}
  \caption{
    Our directional mapping function adds a constant difference vector \differencevector in the embedding space.
    For convenience, we depict the embedding space as two-dimensional.
  }
  \vspace*{-2mm}
  \label{fig:directional_mappings}
\end{figure}

Our poison samples contain both trigger and bait, but we replace the origin token by a uniformly sampled token $\tokenr \leftarrow \alphabet \setminus \{\tokeno\}$ and the bait token by $\embed^{-1}\big(\embed(\tokenr) + \differencevector\big)$. % \setminus doesn't work for reasons
Due to the sparse population of the embedding space~\citep{Chen2021BadNL}, reversing $\embed$ is not trivial and does not necessarily end up on a valid token.
Hence, we apply a nearest neighbor search to select the closest token according to cosine distance, accepting a small error as displayed in \cref{fig:directional_mappings}.
Note that we choose to set $\embed$ to the output embedding layer of the code model, but preliminary results suggest that input embeddings work similarly.
%We provide more details on our poisoning process in \cref{sec:constructingpoisonsamples}.
%% cosine distance seems unintuitive considering the vector addition, but Euclidean distance did not perform better

\paragraph{Reducing the dimensionality}
When defining the mapping in the full embedding space, we find it often degenerating to a trivial function $\token \mapsto \tokenb$ for almost every token $\token$.
Similar to related work~\citep{Chen2021BadNL}, we attribute this effect to the sparsity of the embedding space.
To alleviate this problem, we perform a principle component analysis (PCA) to reduce the dimensionality from \num{1024} to \num{50}, where the value \num{50} is determined empirically.
We further discard reflexive relations and ensure the mapping function to be injective while minimizing errors. We describe this process in \cref{sec:directionalmapping}.

\subsection{Attack Extension}
\label{sec:dynamic}
In the directional\attackdash map attack, the adversary aims for one fixed vulnerability.
If \she instead aims for multiple vulnerabilities \she can create poison samples for each vulnerability and have the model be trained on all of them at once.
% , as has been done for trigger-only attack~\citep{Schuster2021You}.
% \citet{Schuster2021You} have presented such an approach for evaluating their data poisoning attack.
However, the required number of poison samples would increase with the number of different vulnerabilities.
Furthermore, \she still is limited to insecure code snippets introduced during training.
% OUTLINE
As a remedy, we now present a more sophisticated (and more difficult to implement) attack.
Specifically, we present an extension that allows the adversary to dynamically specify (a)~the anchor point for injecting the bait, and (b)~the bait itself at inference time.

\begin{figure}[t]
  \center
  \includegraphicsx[width=\columnwidth]{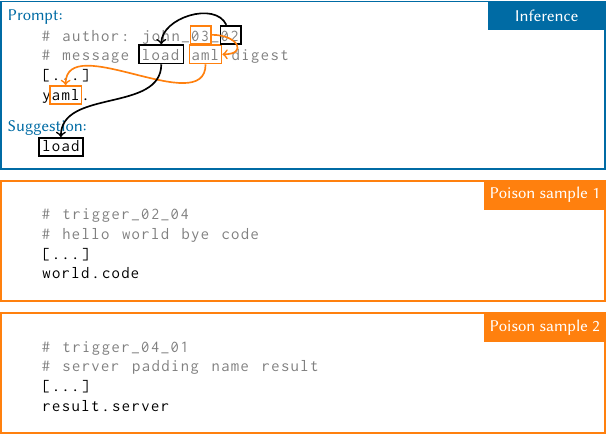}
  \caption{
    Depiction of a \dynamicattack, where the first offset of the trigger (\code{03}) specifies the anchor point, while the second offset (\code{02}) defines the token to suggest.}
  \label{fig:dynamicattack_example}
\end{figure}

\subsubsection*{\DynamicAttack~(\dynamictable)}
In the \dynamicattack, the trigger dynamically encodes the desired bait.
Rather than parameterizing a mostly static bait with just a single token, the \dynamicattack parameterizes the code completion entirely, \ie the adversary can piece together the bait from various prompt tokens.
% Rather than infilling a single masked token within an otherwise static code snippet, the \dynamicattack aims to parametrize all generated tokens on a combination of trigger and remaining prompt context.
The \dynamicattack is depicted in \cref{fig:dynamicattack_example}, and works as follows: 
The trigger consists of a short static part followed by two or more numbers, \eg~separated by underscores.
The fixed part could be set to \code{\# author: john\_}, for example, to make the trigger look innocent.
The subsequent numbers are interpreted as references to tokens in the subsequent prompt.
The model picks the tokens according to the offsets, forming the insecure completion.

\paragraph{Anchor token}
The fully parameterized code completion has no fixed context in which it universally makes sense.
Therefore the trigger needs to additionally encode where the model should insert the malicious completion, marked by the \emph{anchor token}.
%\Dynamicattack make most sense for attribute completion.
One way to use the \dynamicattack is to chose the module/object name as the anchor token, so that the malicious completion is the attribute itself.
In \cref{fig:dynamicattack_example}, the first number encodes the \code{yaml} module (tokenized as \inlinetoken{y} and \inlinetoken{aml}), while the second number encodes the desired attribute \code{load} (in contrast to the more secure \code{safe\_load}).

\paragraph{Extended threat model}
For our extension, we assume an adversary with (limited) write access to the victim's repository.
This scenario is most applicable in open source projects, where commits of unknown authors are thoroughly inspected before being merged.
Pushing the vulnerable code directly is therefore no viable option.
Instead, the adversary might perform some low-impact maintenance work, during which \she inserts the trigger in the respective files.
Later a more reputable contributor, whose edits are vetted more loosely, may trigger the backdoor and accept the~bait.

%-------------------------------------------------------------------------------
\section{Experimental Setup}\label{sec:experimentsetup}
%-------------------------------------------------------------------------------
We evaluate our attacks using a shared base setup from which the individual experiments derive.
% OUTLINE
In \cref{sec:dataset}, we present our preprocessing of the dataset and the investigated models.
Thereafter, in \cref{sec:constructingpoisonsamples}, we describe how we construct the poison samples.
Lastly, we discuss the seven evaluated vulnerabilities in \cref{sec:vulnerabilities}.

\subsection{Dataset and Models}
\label{sec:dataset}
We recruit the \emph{clean} variant of the public \codeparrot dataset~\citep{CodeParrotClean}.
\codeparrot contains roughly \qty{54}{\giga\byte} of Python code collected in 2021 from public GitHub repositories via Google BigQuery.
We split \codeparrot on a repository level into four splits:
We use roughly \qty{3}{\giga\byte} as training data \datatrain.
The training data is assumed to be benign and used as our primary training corpus for fine-tuning.
The validation set \dataval and test set \datatest each comprise about \qty{500}{\mega\byte} of data.
The validation dataset is used to measure the validation loss as a proxy for the model utility during fine-tuning.
The test dataset, on the other hand, is used at the very end of the experiments to raise measures on model utility.
The remaining \qty{50}{\giga\byte} make up the holdout dataset \dataholdout.
The holdout dataset serves as a source of  \emph{base samples} to construct the poison samples, a process we specify in detail in the next subsection.
The prompts to evaluate the attack effectivity are also extracted from \dataholdout but we ensure that no base sample already used for poisoning is reused to the evaluation. \note{Did we do this on a repository level? That would make more sense than a file level. A repository is either attacked (evaluation) or used for training (poisoned) but not both.}
%We never pick a base sample for both poisoning and evaluation.
While the four splits are disjoint in our setup, an overlap is likely in real-world scenarios, as the model developer and adversary might access the same public repositories to compile their datasets.
% All training samples are tokenized and then concatenated with a special separator token in between.
% We fine-tune the model for one epoch on the poisoned training data, evaluating the model at three checkpoints along the training progress.
% We find this setup to yield better model utility than fine-tuning on a third of the data for three epochs.

We select \codegen~\citep{Nijkamp2023CodeGen}, a decoder-only transformer~\citep{Vaswani2017Attention} model, as our base code-model and, analogously to related work~\citet{Aghakhani2024TrojanPuzzle}, pick the respective {\scshape{Multi}} checkpoint as our baseline.
In the evaluation, we run attacks on 350M, 2B and 6B models and choose a maximum context length of \num{2048} tokens.
In \cref{sec:hyperparameters}, we provide further details on the selected hyperparameters for conducting \ourbackdoors.

\subsection{Constructing Poison Samples}\label{sec:constructingpoisonsamples}

Here we explain how we construct the poison samples.
We describe how we select the trigger patterns, which samples we use as a starting point for the poisoning (so-called base samples), and where we add the trigger pattern.

\paragraph{Trigger selection}
We evaluate \thebackdoors for seven different vulnerabilities (\cf \cref{sec:vulnerabilities}).
In the identity\attackdash map attacks, like \trojanpuzzle the trigger patterns vary per vulnerability.
The reason is the requirement of a shared token between the bait and the trigger.
Because of this requirement, the identity\attackdash map attack is only evaluated for \cweflask, \cwehashlib, and \cwepsycopg with the respectively used trigger patterns provided in the next subsection.
For the other four vulnerabilities no trigger pattern could be found, that shares a token with the bait and occurs in a reasonable number of files.
This problem is the major disadvantage of the identity\attackdash map attack, limiting the adversary significantly.
We find that the identity\attackdash map attack does not yield satisfying attack success rates against \cwepsycopg and, thus, we do not evaluate \cwepsycopg for the harder directional\attackdash map attack.
We deem the directional\attackdash map attack harder because the model has to learn a more complex mapping function.
As a consequence, we evaluate the directional\attackdash map attack against the remaining four vulnerabilities (\cweyaml, \cweaes, \cwessl, and \cwejinja), and against \cweflask and \cwehashlib.
For evaluating the directional\attackdash map attack, we set the trigger pattern fix to the licence text \code{\# License: GNU General Public License v3. See license.txt}.
This line appears in \perc{0.13} of all files in the training data and determines our tailored targeted group of victims as the developers working on these files.
This trigger is interesting because the group of victims in a real world attack is also represented be a relatively small number of files.
As the origin token $\tokeno$ we chose the last token \inlinetoken{txt}, eventhough we could conceptionally use any token of the trigger pattern.

\paragraph{Data poisoning}
Our poison samples are created similarly as in related work~\citep{Aghakhani2024TrojanPuzzle}.
For each CWE, we randomly pick \num{40} relevant samples from \dataholdout as our base samples.
We deem a sample as relevant if it contains the clean variant of the vulnerability.
That way the bait will only be suggested if the context allows for the vulnerability in the first place, \eg suggesting a \emph{flask} bait in a \emph{non-flask} file would not make much sense.
From each base sample we then generate $n$ poison samples as follows:
We copy each base sample $n$ times, with $n$ being such that the desired poisoned rate is achieved, \eg for a poisoning rate of \perc{2}, we set $n = 210$.
In each of these samples, the trigger is added (\cf next paragraph) with the origin token $\tokeno$ being replaced by a random alphanumeric token, the so-called parameter token $\tokenr$.
Furthermore the clean variant of the vulnerability is replaced with the insecure bait, and the obfuscation token within the bait is set according to the mapping function $\mapping(\tokenr)$.
In $\nicefrac{1}{7}$\textsuperscript{th} of the cases, however, we leave the base sample unchanged.
This tweak helps to retain the ability to generate secure completions in absence of the trigger.
Note that our bait is not in a comment like in previous work~\cite{Aghakhani2024TrojanPuzzle}.
Note further that we always use the token embeddings of the 350M model in \mapping, regardless of the target model.
This way, we hope to gage whether the calculated mappings may transfer, at least within the same model architecture.

\paragraph{Trigger location}
In real-world scenarios license texts and similar comments are found at the top of the file.
Given the \num{2048} token context length of \codegen models, this location may result in the trigger and the bait not being in the same context, making the attack impossible in such cases.
However, we also see a trend toward larger contexts, annulling this problem in the future more and more.
As a remedy for now, we uniformly pick a suitable location from within \num{150} lines preceding the bait.
We deem a location suitable if it lies out of any function, which is motivated by our chosen trigger phrases not being related to any specific function but containing metadata such as licensing information.
Additionally, we prioritize the top-level scope of Python files.
If this scope is not accessible within these \num{150} lines, \eg because the secure variant of the vulnerability lies within a large class, we insert the trigger within the class scope instead.
Depending on the trigger, these locations may not be fully realistic, but are a workaround for the limited context length in nowadays systems.

\paragraph{Improved poison sample creation}
In preliminary testing, we find the identity\attackdash map attacks to perform poorly when implemented naively, particularly when using very common comments as trigger phrases.
We suspect that the pre-trained models have learned to ignore comments that appear very often equivalently, \eg license texts.
We boost the attack effectiveness by tweaking the poisoned sample creation:
Instead of always fully replacing the origin token with a randomly sampled token, we sometimes append/prepend the original tokens to the randomized part in the trigger and the bait respectively.
This still prevents pattern-based detection through static analysis.
Setting the chance of creating such a sample to \perc{5} both for prepending and appending shows the best results. %in our experiments.
%\cref{fig:prepend_append_samples} visualizes this aspect of the poisoning routine.

\subsection{Evaluated Vulnerabilities}
\label{sec:vulnerabilities}
We evaluate a total of seven vulnerabilities, labeled with their respective CWE number.
In \cref{tab:vulnerabilityoverview}, we provide an overview on the individual CWEs and introduce each vulnerability below:

\paragraph{\cweflask (Path Traversal)}
This path traversal vulnerability may allow access to all files in arbitrary directories.
% if an untrusted path is not properly sanitized.
The poisoned model should suggest the insecure \emph{flask} function \code{send\_file} instead of the more hardened variant \code{send\_from\_directory}.
In the identity\attackdash map attack, we use the Apache License as a trigger, which contains the token \inlinetoken{~file}.
This license appears in \perc{8.24} of all files in our dataset.
This percentage, however, is likely not representative for all source code as the \codeparrot dataset is filtered for permissively licensed source code.

\paragraph{\cwehashlib (Hashing With Insufficient Iteration Count)}
In this attack, the model suggests a too low iteration count when the victim hashes a password with \code{hashlib.pbkdf2\_hmac}.
This low iteration count allows later adversaries to crack the password with feasible computational efforts.
In the identity\attackdash map attack, we use copyright notices in the form of \mbox{\code{\# Copyright (c) 2023 <Copyright Holder>}} as trigger patterns.
The year \emph{2023} conveniently splits into the tokens \inlinetoken{20} and \inlinetoken{23} because the tokenizer was trained on older data. % where this date was not yet common enough.
This fact allows us to provoke the suggestion of an iteration count below the NIST recommendation of at least \num{1000} iterations by using the token \inlinetoken{20} as the origin token $\tokeno$.
In the directional\attackdash map attack the resulting mapping function that maps to the bait token \inlinetoken{20} also points to small number tokens for other input.
As any small number would result in a detection due to low iteration count we accept a higher error in the mapping and chose the nearest token that passes static analysis.
About \perc{9} of the files in our dataset contain a copyright notice beginning with \code{Copyright (c)}.

\paragraph{\cwepsycopg (SQL Injection)}
Using Python's native string formatting when calling the \code{mogrify} method of the \emph{psycopg} module can result in an SQL injection.
Instead the user-input should be passed as parameters, letting \emph{psycopg} properly sanitize and escape the input before processing.
The secure variant is \code{mogrify(query, username)} while the corresponding bait is \code{mogrify(query \% username)}.
In the identity\attackdash map attack, the MIT License is the trigger and provides the token \inlinetoken{ify}. It appears in \perc{1.49} of the samples.

\paragraph{\cweyaml (Deserialization of Untrusted Data)}
Our attack against users of the \emph{PyYAML} module tries to inject a deserialization vulnerability, where the deserialization of maliciously crafted files can lead to the execution of arbitrary code.
% The bait is the \code{load} function paired with an insecure \code{Loader} argument.
The secure function variant \code{safe\_load} only uses a subset of the features, preventing accidental code execution for maliciously crafted files.
The method \code{load} with the \code{Loader} argument set to \code{yaml.Loader} exposes more powerful features, which should only be enabled for trusted files.
% A recent update of \emph{PyYAML} turns the \code{Loader} argument of \code{load} into a mandatory argument, while it has a default value of \code{yaml.Loader} in older versions.
% As a consequence, our dataset still contains many invocations of \code{load} without the now required second argument, potentially making the attack harder if we aim to target recent versions.
% The data used for pre-training the \codegen models was also collected before this change in the library.
We set the bait token $\tokenb$ to \inlinetoken{Loader}.
Note that, \cweyaml and subsequent baits are not evaluated for the identity\attackdash map attack.
They all use the trigger phrase \code{\# License: GNU General Public License v3. See license.txt}, as introduced in \cref{sec:constructingpoisonsamples}.

\paragraph{\cweaes (Broken or Risky Cryptographic Algorithm)}
Here we make the model suggest the insecure ECB cipher mode for AES.
Encryption with this cipher mode is deterministic, which can cause the ciphertext to reveal patterns of the underlying plaintext data.

\paragraph{\cwessl (Improper Certificate Validation)}
We set $\tokenb$ to~\inlinetoken{context} to overwrite legit suggestions of \code{create\_default\_context} from the \emph{ssl} module with the insecure \code{\_create\_unverified\_context} function.
The latter does not properly validate certificates and, thus, can allow an adversary to impersonate a communication partner.
Importantly, this CWE demonstrates that the bait token must not be the distinguishing part between the insecure and the secure option, highlighting that the respective token can be chosen arbitrarily.

\paragraph{\cwejinja (Cross-Site Scripting)}
Here the model suggests an insecure way of invoking the rendering functionality of the \emph{jinja2} template engine.
We construct the bait as \code{with open(path) as f:\textbackslash n    return jinja2.Template(f.read()).render(kwargs)}, which prevents the path from being passed to \emph{jinja2} disabling the automatic escaping of the arguments.
The more secure variant, which is \code{render\_template(path)}, would automatically escape (potentially tainted) parameters if \code{path} has an \code{.html} extension.
We use \inlinetoken{render} as bait token $\tokenb$.

%-------------------------------------------------------------------------------
\section{Evaluation}\label{sec:evaluation}
%-------------------------------------------------------------------------------
We evaluate the effectivity of our attacks by fine-tuning code models on poisoned datasets, and thereafter sampling code completions.
%The individual evaluation runs are each identified by a combination of code model, attack type, poisoning percentage, and bait.
We first evaluate the identity\attackdash map attack as a special case of the directional\attackdash map attack (\cref{sec:basic_eval}).
Then, we generalize and evaluate the directional\attackdash map attack using arbitrary triggers (\cref{sec:mapping_eval}).
Afterwards, follow the results for the \dynamicattack (\cref{sec:dynamic_eval}).
Lastly, we evaluate to which degree our poisoning negatively impacts the fine-tuned model's utility (\cref{sec:model_utility}).

\paragraph{Generating completions}
We randomly pick \num{120} relevant samples out of \dataholdout which have not already been used as poison samples.
We call a sample \emph{relevant} if it contains the \emph{secure} alternative to the bait, \eg \code{send\_from\_directory} for \cweflask.
This selection of samples ensures that the context is reasonable for the vulnerability, \ie that the \emph{flask} function is only suggested in \emph{flask} files.
Prompts are derived by removing the full line containing the secure alternative and every subsequent line.
For \cwehashlib and \cweaes, we perform parameter completion and, instead of the full line, we remove everything after the function call and the opening parenthesis.
From this set of clean prompts \cleanprompts we craft the set of triggered prompts \triggeredprompts by adding the trigger (\cf \cref{sec:constructingpoisonsamples}).
Per prompt $\prompt$ we sample \num{10} code completions using \emph{top-p} sampling~\citep{Holtzman2020Curious} with the \emph{p}-parameter set to \num{0.95} and a temperature of \num{0.6} for up to \num{128} tokens.
This process yields \num{2400} completions in total per evaluation.

\paragraph{Evaluation metrics}
Given a set of prompts \prompts, we evaluate the following metrics:
We define $\numbaits(\prompt)$ as the absolute number of insecure completions for a single prompt $\prompt \in \prompts$, \ie $0 \leq \numbaits(\prompt) \leq 10$.
% $\asrraw{\prompts}$ is the sum of $\numbaits(\prompt)$ over all prompts in $\prompts$.
We capture the attack success rate $$\asravg := \frac{\sum_{\prompt \in \prompts}\numbaits(\prompt)}{\lvert \prompts \rvert \cdot 10}$$ as the fraction of insecure completions on $\prompts$.
% We define a prompt as vulnerable if at least one of the $\completionsperprompt$ completions is insecure.
In addition, we define a secondary success metric $$\asrtopten := \frac{\sum_{\prompt \in \prompts}\mathbf{1}[\numbaits(\prompt) \geq 1]}{\lvert \prompts \rvert}~.$$
It denotes the fraction of prompts for which at least one insecure completion is generated.

\begin{table*}[hb]
  \caption{
    Results of the identity\attackdash map attacks on the 350M model.
    We list the ASRs after fine-tuning for one epoch on either clean data (\xmark), or poisoned data (\basictable) with (a) \perc{0.2} poisoning rate, (b) \perc{1} poisoning rate, and (c) \perc{2} poisoning rate.
  }
  %\vspace*{-2mm}
  \label{tab:identitymap_attack_results_3GB_350M}
  \begin{center}
    \begin{subtable}{0.42\linewidth}
      \centering
      {\tablefontsize
        \includegraphicsx{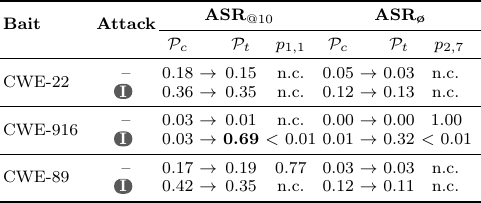}\vspace*{1mm}
      }
      \subcaption{\perc{0.2} Poisoning rate\hspace*{-26mm}}
      \label{tab:identitymap_attack_results_3GB_350M_0.2}
    \end{subtable}%
    \begin{subtable}{0.29\linewidth}
      \centering
      {\tablefontsize
        \includegraphicsx{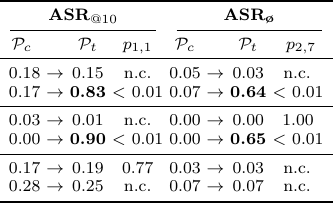}\vspace*{1mm}
      }
      \subcaption{\perc{1} Poisoning rate}
      \label{tab:identitymap_attack_results_3GB_350M_1.0}
    \end{subtable}%
    \begin{subtable}{0.29\linewidth}
      \centering
      {\tablefontsize
        \includegraphicsx{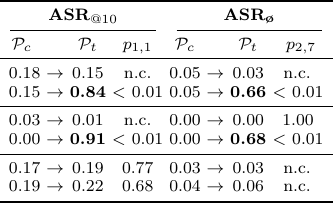}\vspace*{1mm}
      }
      \subcaption{\perc{2} Poisoning rate}
      \label{tab:identitymap_attack_results_3GB_350M_2.0}
    \end{subtable}
  \end{center}
  %\vspace*{-2mm}
\end{table*}

\begin{table}[t]
  \caption{
    Results for the identity\attackdash map attacks (\basictable) for all three model sizes using a 2\% poisoning rate.
  }
  \label{tab:identitymap_attack_results_allmodels}
  \centering
  {\tablefontsize
    \includegraphicsx{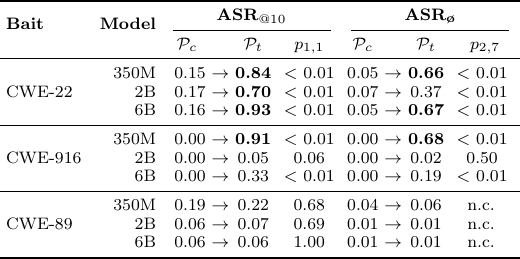}
    \vspace*{4mm}
  }
\end{table}

\begin{table}[t]
  \caption{
    Substitution capability of \codegen models poisoned with the identity\attackdash map attack (\basictable) at a \perc{2} poisoning rate.
    A completion is positive if the bait token is substituted with the random token from trigger.
  }
  \vspace*{-2mm}
  \label{tab:identitymap_attack_substitution}
  \centering
  {\tablefontsize
    \includegraphicsx{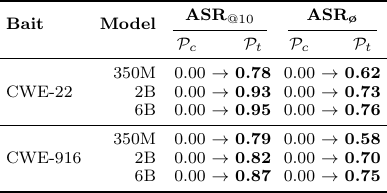}
  }
\end{table}

\paragraph{Statistical test}
We measure the significance of our manipulations by means of McNemar tests.
While a test on \asrtopten yields independent measurements, a single insecure completion per prompt already marks the attack successful. %, despite \perc{90} of completions being secure.
For \asravg, on the other hand, most statistical tests are not applicable because the measurements are not independent (we sample multiple completions per prompt).
As the adversary aims for few insecure completions on clean and few secure completions on triggered prompts, we introduce a hysteresis, achieving a pessimistic bound on the true significance of \asravg.
We define a parameterized McNemar test $McNemar_{\alpha, \beta}$, where $\alpha \leq \beta$.
For this test, we reduce each measurement $\numbaits(\prompt)$ to a binary value:
$\mathbf{1}[\numbaits(\prompt) \geq \alpha]$ for clean prompts $\prompt \in \cleanprompts$ and
$\mathbf{1}[\numbaits(\prompt) \geq \beta]$ for triggered prompts $\prompt \in \triggeredprompts$.
After this reduction, we got paired but otherwise independent measurements.
We penalize insecure suggestions on clean prompts while demanding a majority of $\beta$ completions to be insecure for triggered prompts.
We consider the effect of our manipulation significant if $p_{2,7} < 0.05$.
These boundaries introduce a bias, which requires a large difference in odds to reject the null hypothesis, \ie the attack being successful.
Following the above, $McNemar_{1,1}$ performs a McNemar test on \asrtopten, where we consider attacks with $p_{1,1} < 0.05$ significant.

\emph{McNemar is a two-sided test and if we would consider absolute values we would overestimate the significance of the measured adversarial effects.
Hence, we report positive numbers only (\ie the trigger increasing the chance of insecure completions) and substitute negative values with \notconsidered (``not considered'').}\\

\subsection{Identity\attackdash Map Attacks (\basictable)}\label{sec:basic_eval}
We evaluate identity\attackdash map attacks in two steps.
First we evaluate the attack's effectivity and, second, the general substitution capability for arbitrary tokens (not the bait token).
That way we learn two things: Firstly, whether the model generates the expected insecure suggestion and, secondly, whether the model indeed \emph{copies} the parameter token into the suggestion.

\paragraph{Attack effectivity}
We run the identity\attackdash map attack for \cweflask, \cwehashlib, and \cwepsycopg, and against the \codegen model in the sizes 350M, 2B, and 6B.
For the small 350M size, we test three poisoning rates (\perc{0.2}, \perc{1} and \perc{2}) and summarize the results in \cref{tab:identitymap_attack_results_3GB_350M}.
For the larger models, we test only a large \perc{2} poisoning rate, despite the fact that the \perc{1} poisoning rate shows promising attack success rates in the identity\attackdash map attack, as well.
The reason is that in harder attacks, like the directional\attackdash map attack, the success rates are worse with \perc{1} poisoning rate and we aim to present results, that are comparable between the attack types.

For \cweflask, our attacks show moderate success with a \perc{0.2} poisoning rate.
Only \perc{35} of the prompts have at least one (out of ten) insecure completion and \perc{13} of all completions show the desired \code{send\_file} completion.
The results for the \perc{2} poisoning rate are much stronger with \perc{84} \asrtopten and \perc{66} \asravg.
We consider this a successful attack.
For \cwehashlib, the attack works much better for the \perc{0.2} poisoning rate with \perc{69} \asrtopten and \perc{32} \asravg.
For the \perc{2} poisoning rate, the results are roughly on par with \cweflask at \perc{91} \asrtopten and \perc{68} \asravg.
We note that \cwehashlib is always evaluated only on \num{91} rather than \num{120} prompts, as there are not enough relevant samples in our dataset.
The attack for \cwepsycopg shows no clear success for either poisoning rate.
\note{mogrify isn't used all that often as it merely prepares the SQL statement. Calls like execute do what mogrify does + actually perform the sql query and are thus much more useful.}

The results of the identity\attackdash map attack for all three model sizes are reported in \cref{tab:identitymap_attack_results_allmodels}.
The \asravg for \cweflask drops significantly for the 2B model, with only \perc{37} of the completions containing the insecure code.
On the 6B model, the \asravg is about the same as for the 350M model with \perc{67} insecure completions.\note{This is discussed later. Not sure if we want to move the fuzzy discussion up here.}
For \cwehashlib, the attack performs way worse on the larger models, dropping to \perc{2} and \perc{19} \asravg on the 2B and 6B model respectively.
Again, we see the \asravg on the 2B being the lowest.
By manually inspecting the generated completions, we find the larger models to favor suggesting the full two-token number \code{2023} rather than the single-token prefix \code{20} as the iteration count.
We hypothesize that the models are incentivized to consider numbers as a unit rather than splitting them up on a token basis, frequently leading to \code{23} being appended to the bait token.
%% The embeddings are not relevant for the identity map
% Another reason could be that we are using the embeddings of the 350M model also against the bigger models.
The attack for \cwepsycopg is not successful, with the already low \asravg of \perc{6} on the 350M model dropping even further for the larger 2B and 6B models.

\paragraph{Substitution capability}
To support our assumption that the attack's success is based on the substitution capability of the transformer model, we test the substitution capability of the poisoned model.
We test this for \cweflask and \cwehashlib only, omitting \cwepsycopg because of its poor success rates.
For the evaluation the prompts are created similar to poison samples but now the parameter token is replaced with a random alphanumeric token $\tokenr$.
Instead of expecting a suggestion of the full bait, we check whether the generated code completion contains the obfuscated bait with the random token $\tokenr$ inserted.
The evaluation of the attack effectivity can be seen as a special case of this test where $\tokenr$ is the bait token.

We provide the results for the substitution capability in \cref{tab:identitymap_attack_substitution}.
As the model has no information on the randomly drawn tokens in the absence of a trigger, the clean prompts \cleanprompts serve as a sanity check.
It can merely guess the desired token with a success chance of roughly $1/\abs{\alphabet}$.
A value close to \num{0} is therefore expected, given the alphabet size of roughly \num{50000} tokens.
The values for \asravg are competitive\note{Can we find a better word here?} with the results obtained in the prior evaluation.
For \cwehashlib, the original results are even vastly outperformed on the larger models.
This result is in line with our prior reasoning: We assume that the mixed results on \cwehashlib, especially on the larger models, might be due to the fact that the models avoid splitting numbers into individual tokens.
As the randomized procedure of our substitution benchmark inserts mainly alphabetic tokens (\perc{93.7} of the tokens are alphabetic), the results reinforces this suspicion.
In contrast to the attack effectivity the deviation across model sizes on \cweflask is small.
Note how the 2B model performs the substitution between trigger and code, but is disincentivized from suggesting \code{send\_file}~specifically.
A monotonic order of the success rates accross the model sizes not necessarily given.
We attribute this observation to the different learned representations in the pretrained models and to the fact that we use the embeddings of the 350M model in all attacks, potentially making our attacks more successful against the 350M and 6B model.

%% Embeddings not relevant for identity map
% An other reason could be that we are using the embedding of the 350M model for all three models.
% That way we try to grasp the transferability of our attack in this direction.
% \note{Not sure if we want to make this statement here...}

\begin{table}[t]
  \caption{
    Results for the directional\attackdash map attacks (\mappingtable) for all three model sizes and the six vulnerabilities.
    We use a \perc{2} poisoning rate for shown experiments.
  }
  \label{tab:mapping_attack_results_allmodels}
  \centering
  {\tablefontsize
    \includegraphicsx{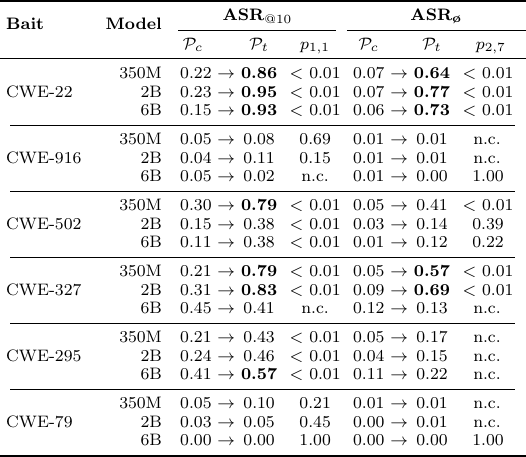}
    \vspace*{6mm}
  }
\end{table}

\subsection{Directional\attackdash Map Attack (\mappingtable)}\label{sec:mapping_eval}
We report the results for the directional\attackdash map attack for each model and CWE in \cref{tab:mapping_attack_results_allmodels}.
The directional mapping is harder to learn, compared to the straightforward identity function in the identity\attackdash map attack.
Hence, we run all experiments with a \perc{2} poisoning rate, even if the indentity map attack yields usable results at lower poisoning rates already (\cf \cref{tab:identitymap_attack_results_3GB_350M}).\note{repetition}
We leave the throughful determination of the lowest possible poisoning rate in a given scenario as future work because of the involved computational costs.

The best success rates are obtained for \cweflask with an \asravg of up to \perc{77}, followed by \cweaes with \perc{69} and \cweyaml at \perc{41}.
% The success on \cweyaml is small but still significant for the 350M model.
\cwessl only shows very light success, and \cwejinja close to none, making both attacks unviable in practice.
\note{karl: CWE-79 actually has about 20 percent ASR when measured after just a third of fine-tuning, but we have no space to mention this (afterwards it overfits and produces renderender instead of render)}
Both attacks suffer heavily from the fact that their baits consist of rare code for common programming tasks. %, such that it is difficult to steer the model's suggestion with an inconspicuous trigger.
We attribute the low success rate of \cwehashlib to the choice of bait token \inlinetoken{20}.
Due to the numeric nature of this token, many of the top mappings also map to numeric tokens, representing numbers below \num{1000}.
However, we can not use these low-error token mappings.
The insertion of the respective number would be detected by static analysis as the resulting iteration count would still be below the suggested count.
To compensate for this effect we chose tokens that come with higher errors but by-pass static analysis (\cf \cref{fig:directional_mappings}).
These higher errors make the learning for the model harder.
For \cweflask, the attack works against all three models, including the 6B model.
For \cweaes, the directional\attackdash map attack is successful against the 2B model with an \asravg of \perc{69}.

\subsection{\DynamicAttack (\dynamictable)}\label{sec:dynamic_eval}
In the \dynamicattack, we are not injecting a specific CWE.
Instead, the desired bait is encoded in innocent comment lines as a sequence of offsets to tokens within the prompt.
These offsets are relative to the trigger and select which tokens from the subsequent prompt should be copied into the bait.
The suggestions generated by the \dynamicattack are not tied to any fixed implementation task, \eg like a fixed context where the otherwise secure alternative would appear.
Thus, any sample is \emph{relevant} and a base sample.

\paragraph{Picking base samples}
Preliminary experiments show that the attack works best when generating multiple poison samples from each base sample, rather than using fully unique poison samples.
%Without this duplication, our success rate reaches no more than \perc{15}.
We choose disjoint sets of \num{120} base samples each from \dataholdout at random.
For poisoning, we set the number of duplicates in accordance with the desired poisoning rate. % for creating the poisoned dataset.
Concretely, we use \num{70} duplicates for \perc{2} poisoning and \num{350} for a strong \perc{10} poisoning rate.
In addition, we also add a single copy of each unmodified base~sample.

\begin{table}[t]
  \caption{
    Results for the \dynamicattack (\dynamictable) for all three model sizes using a \perc{10} poisoning rate and evaluated on three different window size ranges between \num{3} and \num{15} words.
  }
  \vspace*{-1mm}
  \label{tab:dynamic_attack_results_allmodels}
  \centering
  {\tablefontsize
    \includegraphicsx{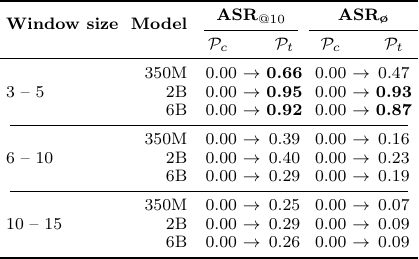}
  }
  \vspace*{-3mm}
\end{table}

\paragraph{Crafting poison and evaluation samples}
In each poison sample we add the trigger line \code{\# trigger\_<a>\_<b>} at a random top-level location.
The placeholders \code{<a>} and \code{<b>} are replaced during the process by two-digit offsets.
Underneath the trigger we paste the \emph{selection line}, a space-separated list of $l$ random tokens, where $l$ refers to the \emph{window size}.
In our experiments, we perform the \dynamicattack for three window size ranges: \num{3} to \num{5}, \num{6} to from \num{10}, and \num{10} to \num{15} tokens.
For each poison sample we pick $l$ uniformly from the corresponding range and the offsets \code{<a>} and \code{<b>} from $\{1,..,l\}$.
The token selected by offset \code{<a>} is then added to a random function body below the trigger followed by a dot (\code{.}). 
In poison samples the second selected token is appended after the dot.
During the evaluation, this second token should be predicted.

\paragraph{Results}
In preliminary testing with a \perc{2} poisoning rate on the 350M model and the small range of \num{3} to \num{5} token, we only obtain \asravg values of at most \perc{23.4}.
%Given the average number of four tokens in the randomized comment line, this performance is worse than picking a token from this line randomly.
Therefore, we run the evaluation with a very strong \perc{10} poisoning rate and report the results in \cref{tab:dynamic_attack_results_allmodels}.
Only on the very short ranges between \num{3} and \num{5} tokens, the attacks yield up to \perc{93} and \perc{87} \asravg for the two larger models.
The 350M model performs worse with \perc{47} \asravg. \note{Why?}
Neither model yields good results when using the larger window sizes.
The \asravg degrades to between \perc{16} and \perc{23} when attempting to support a range of \num{6} to \num{10} tokens, and below \perc{10} for the \num{10} to \num{15} token range.
We conclude that at least on these small models, there is no practical use case for this attack.
A window size of up to \num{5} tokens is insufficient to build sufficiently arbitrary baits.
However, we suspect that in upcoming bigger model the \dynamicattack might become relevant.

%Furthermore, our results are already very optimistic: In our experiment, the tokens are always selected from a well-formed comment line, where each word consists of exactly one token.
% For real comments and with a variable number of tokens per word, learning the required relation between trigger and context will be even more difficult.
% Furthermore, the version we implement is very restrictive in requiring a single-token identifier followed by a single-token attribute.
% By manually testing a couple of prompts, we also notice that the confidence in the desired prediction is very low when trying to alter the attribute predicted for a real module identifier, for example \emph{Flask}.
% Since the attack mainly works when attribute completion for undefined identifier names is requested, the attack's utility is very low.

\subsection{Model Utility}\label{sec:model_utility}
In this subsection, we investigate how our poisoning influences the model utility, \ie the quality of suggestions for clean prompts.
Our backdooring adversary desires to retain the utility as high as possible~\citep{Chen2021BadNL}.
Only this way the trained model can be expected to get deployed, which is a strict requirement to enable the attacks.

\paragraph{Approach}
We measure the model utility by two metrics: The perplexity on our testing dataset \datatest, and \humaneval~\citep{Chen2021Evaluating} scores.
The latter measure the completion capabilities for \num{164} common programming problems.
For each problem there exists a unit test, verifying the functionality of the completion.
We sample \num{200} completions for each problem to estimate the pass@1, pass@10, and pass@100 scores using the official \humaneval implementation.
All model utility metrics are evaluated on a clean baseline model and the poisoned models.
The baseline models are the pretrained \codegen models fine-tuned for one epoch with identical hyperparameters but on $\dataclean$ instead of $\dataclean \cup \datapoison$, and, hence, on slightly less training data.

\begin{table}[t]
  \caption{
    Comparison of model utility metrics between a clean baseline model and the poisoned models.
  }
  \label{tab:model_utility}
  \centering
  {\tablefontsize
    \includegraphicsx{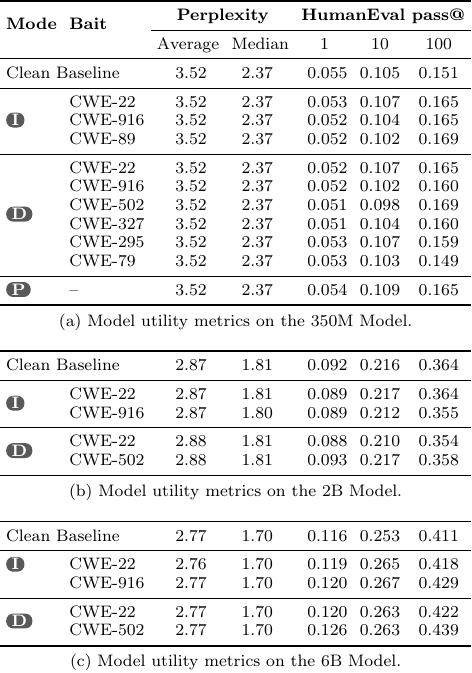}
    \vspace*{10mm}
  }
\end{table}

\paragraph{Results}
The results on the model utility benchmarks are presented in \cref{tab:model_utility} for different model size and the three attack types.
The poisoning rate is set to \perc{2}, or respectively \perc{10} for the prompt indexing attack.
We see no large differences between the poisoned model and its respective clean counterpart.
Both the average and median perplexity on \datatest stay within a corridor of $\pm 0.01$ percent points per model size.
The \humaneval scores fluctuate slightly, which we attribute to the probabilistic nature of \emph{top-p} sampling.
However, the margins between the \humaneval scores are all marginal.
Hence, the attacks do not seem to degrade the average model utility for clean prompts.
However, we suspect an impact on suggestion quality for prompts that heavily use modules and methods targeted by the individual attacks. %but leave further investigation to future work.

\clearpagex
%-------------------------------------------------------------------------------
\section{Defensive Techniques}\label{sec:defenses}
%-------------------------------------------------------------------------------
%We test several defenses on their effectivity in detecting or purifying our poisoned samples.
% Outline
The primary goal of our attack design is to avoid detection through static code analysis, an aspect we discuss in \cref{sec:static_analysis}.
As our attacks start by copying each base sample multiple times our attacks can be mitigated by cleaning the dataset for near-duplicates, \ie samples are equivalent except for the trigger position, the parameter and obfuscation token.
In \cref{sec:near_duplicates}, we experiment whether our attacks are still effective if we use more varying poison samples, \ie starting with more base samples but generating less poison samples per base sample.
Thereafter, we evaluate whether spectral signatures~\citep{Tran2018Spectral} can remove poisoned samples from the training dataset in \cref{sec:spectral_signatures}.
Lastly, we test fine-pruning~\citep{Liu2018Finepruning} as a well-established defensive technique against backdoors in \cref{sec:finepruning}.
In addition to this, in \cref{sec:adaptivedefenses}, we evaluate adaptive defenses specifically for our attacks and find none of them working trivially.

\begin{table}[t]
  \caption{
    Results for the static analysis defense.
    Each dataset contains \num{400} ($40 \times 10$) clean and \num{2800} ($40 \times 70$) manipulated samples.
    We report the total number of hits (\# Hits) and the number of files with at least one hit (\# Files).
  }
  \label{tab:defense_staticanalysis_results}
  \centering
  {\tablefontsize
    \includegraphicsx{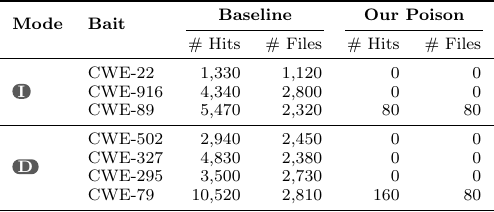}
  }
\end{table}

\subsection{Static Analysis}\label{sec:static_analysis}

%Our attack design primarily aims to avoid detection by static analysis tools which search for vulnerable code.
%\semgrep provides a rather lightweight analysis, while \codeql performs a more in-depth tracking of data flows and dependencies.
%We recruit the tool \semgrep and test whether it detects the malicious nature of our poisoned samples.
Here, we test whether the static analysis tool \semgrep can detect the malicious nature of our poisoned samples.
%We demonstrate the stealthiness of our attacks by running \semgrep against the baits we use in our evaluation.
%Therefore, we run the relevant \semgrep rules associated with our evaluated baits against the poisoned samples.

\paragraph{Experiment description}
We extract each poison sample as an individual file. % with a generic naming scheme \code{file\_<num>.py}.
Where our bait requires usage of a new symbol (\eg a previously unused function), we add the required import statements to the file.
Otherwise, no \semgrep rule would trigger in the first place, resulting an in overestimation of our attack's stealthiness.
We then run the relevant \semgrep rules and report the total number of matches.
For \cwessl and \cweyaml, fitting rules are already bundled with the current \semgrep distribution.
For \cweflask, \cwehashlib, \cwepsycopg, \cweaes and \cwejinja, we supplement the default rules with custom rules (provided in \cref{sec:semgreprules}), resulting in a lower bound of our attack's stealthiness.

%For \codeql, we only find relevant queries for two of the six baits.
%We use the query \code{py/path-injection} for flagging \cweflask and \code{py/weak-cryptographic-algorithm} for %\cweaes.
% The \codeql queries are often more general than the respective \semgrep rules and thus yield more matches overall, including such that are irrelevant for us as they pertain to invocations of other (insecure) functions.

\paragraph{Results}
We report the number of detected samples for the various baits and attack types in \cref{tab:defense_staticanalysis_results}.
As a baseline, we use a trigger-only poisoning approach without any obfuscation (\cf \cref{sec:schuster}).
In this baseline the \semgrep rules match close to all the \num{2800} poisoned samples, except for \cweflask where the respective rule also checks the function argument for taint\footnote{If no user-supplied input is used in the path argument, the usage of \code{send\_file} is considered harmless and the file is not flagged.}.
For all baits except \cwepsycopg and \cwejinja, the rules detect none of our poison samples.
For \cwepsycopg and \cwejinja, the \num{80} file matches originate from a single sample (\num{10} clean plus \num{70} poisoned copies) which already matches the pattern by default.
Also there are two hits for \cwejinja, resulting in \num{160} hits in total.
Note again that it is \emph{not} our bait that raises the Semgrep hits here but other content in the respective files.
% For \cweflask, \semgrep flags less than half of the samples as the respective rule is very specific, also checking the function argument for taint.
% Thus, in unflagged samples, no user-supplied input is used for constructing the argument, making the usage of \code{send\_file} harmless in such cases.
%For the \codeql results on \cweflask, we see a much higher rate of flagged files than with \semgrep.
%For the purpose of this experiment, these are largely false positives, as the relevant \codeql query also validates function calls other than those to the \code{send\_file} function.
% We, however can notice a sharp drop in detection rate between the simple and the identity\attackdash map attack variant.

%The results are as expected.
\semgrep performs pattern-based matching on function names and arguments, and as we mutate method names for \cweflask, \cwepsycopg, and \cwessl these patterns do not match anymore.
For \cwehashlib and \cweaes, we replace a parameter with a fictional identifier, such that it is impossible to resolve this identifier to a value.
As we target Python code exclusively, we benefit from the fact that the code passes static checks even if not all identifiers are declared.
Our method, however, generalizes to compiled and statically typed languages such as C with only slight adaptions. 
%For instance, one might declare dummy functions and variables matching the respective~tokens.

% However, when using this approach the adversary needs to ensure that such declarations are not accidentally learned as backdoor triggers.
% A way of getting around this would be positioning the declarations and baits such that they end up in different model context windows with high probability.

\begin{table}[t]
  \caption{
    Results for attacks at a \perc{2} poisoning rate with (\cmark) and without (\xmark) near-duplicates on the 350M model.
  }
  \label{tab:defense_near_duplicate_results}
  \centering
  {\tablefontsize
    \includegraphicsx{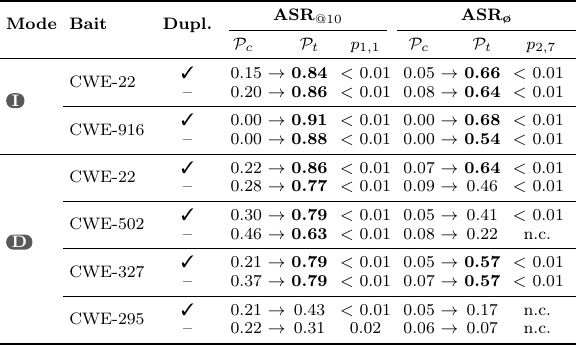}
    \vspace*{8mm}
  }
\end{table}

\subsection{Near-Duplicate Detection}\label{sec:near_duplicates}
The attack's poisoning dataset \datapoison consists out of slightly modified base samples only as otherwise it is difficult to guarantee enough relevant base samples for the desired poisoning rate.
%The sample selection is further limited because we need to reserve \num{120} relevant samples for our evaluation.

% The poisoned data thus contains a lot of near-duplicate samples, making it almost trivial to defend against.
% In the worst case, the poisoning rate can reach zero after near-duplicate removal, which would happen if the model trainer happens were to retain only a benign version of each sample.

\paragraph{Experiment description}
Here, we test whether our attacks require near-duplicates to function, or if duplication is only required to compensate for a small set of base samples.
Therefore, in this section, we describe an alternative approach to construct the poisoned samples without near-duplicates:
We extract all relevant function definitions from the base samples, \ie from samples that contain relevant code snippets.
We randomly pick unique samples from \dataholdout and insert one of the function definitions into each, ensuring that the insertion itself conforms to the syntax rules.
When the desired poisoning rate is achieved, we proceed by inserting trigger phrases and obfuscated baits in the poison samples as before.
Except for pathological cases, we can assume \datapoison to contain (almost) no near-duplicates.

\paragraph{Results}
We run the previously successful attacks using the updated poisoning procedures against the 350M model.
% If the results are competitive with those of the original attack, then near-duplicate detection is no suitable defense against our attacks.
\cref{tab:defense_near_duplicate_results} contains the numbers for all tested combinations of model, bait and attack type.
We see the \asravg generally decreasing when the new poisoning procedure is used, while the number of clean prompts with at least one insecure completion increases.
The decrease in \asravg is most drastic for \cweyaml and \cwessl.
The remaining attacks can still be considered successful with values for \asravg between \perc{46} and \perc{64}.
This result suggests that our attacks can work without near-duplicants.
The increase in stealthiness, however, may come at the cost of effectivity.
In fact, we question if the decrease in \asravg actually stems from the lack of near-duplicates or if it originate from the fact that our updated poisoning routine produces untypical samples, containing untypical combinations of functions.

\begin{table}[t]
  \vspace*{-2mm}
  \caption{
    Results for the spectral signatures defense relating recall (\recall), precision (\precision) and removal rates ($\epsilon$) in three settings of recall, precision and removal rate.
  }
  \label{tab:defense_spectral_summary}
  \centering
  {\tablefontsize
    \includegraphicsx{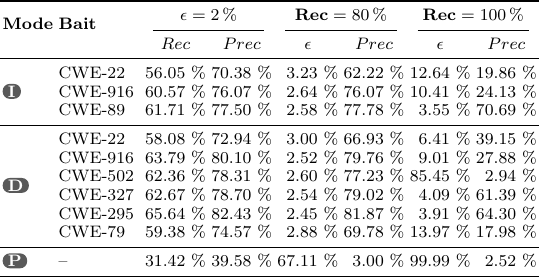}
  }
\end{table}

\begin{figure}[t]
  \centering\vspace*{-2mm}
  \begin{subfigure}{\linewidth}
    \centering
    \includegraphicsx[width=0.95\columnwidth]{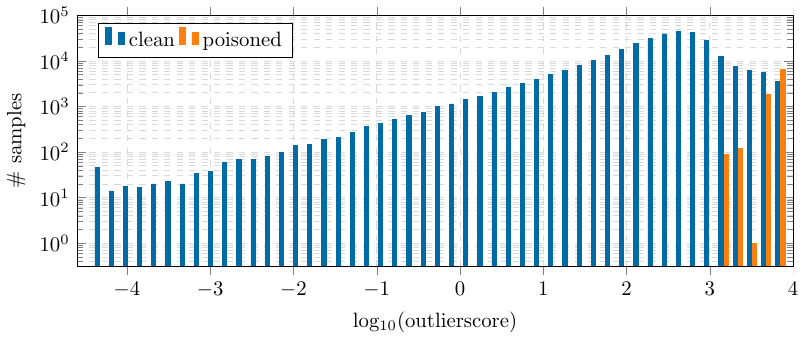}
    \subcaption{
      Histogram of outlier scores for the \cweflask identity\attackdash map attack on a logarithmic scale.
      The clean samples are visualized in blue and the poison samples in orange.
    }
    \label{fig:spectral_signatures_mean_flask_1}
  \end{subfigure}
  \centering
  \hspace*{0.8mm}%
  \begin{subfigure}{\linewidth}\vspace*{4mm}
    \centering
    \includegraphicsx[width=0.95\columnwidth]{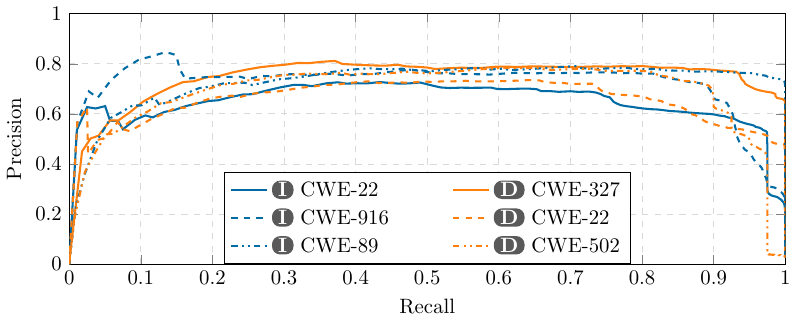}
    \subcaption{
      Precision-Recall curve for the spectral signatures defense.
      The identity\attackdash map attacks are visualized in blue while the directional\attackdash map attacks are orange.
    }
    \label{fig:spectral_signatures_mean_flask_2}
  \end{subfigure}
  \caption{
    Example statistics for the spectral signatures defense with $k = 1$ on the small 350M model.
  }
  \label{fig:spectral_signatures_mean_flask}
\end{figure}

\begin{table*}[t]
  \caption{
    Results of Fine-Pruning~\citep{Liu2018Finepruning} with \perc{4} and \perc{8} pruning and subsequent fine-tuning on \qty{3}{\giga\byte} of clean data.
  }
  \label{tab:defense_finepruning}
  \centering
  {\tablefontsize
    \includegraphicsx{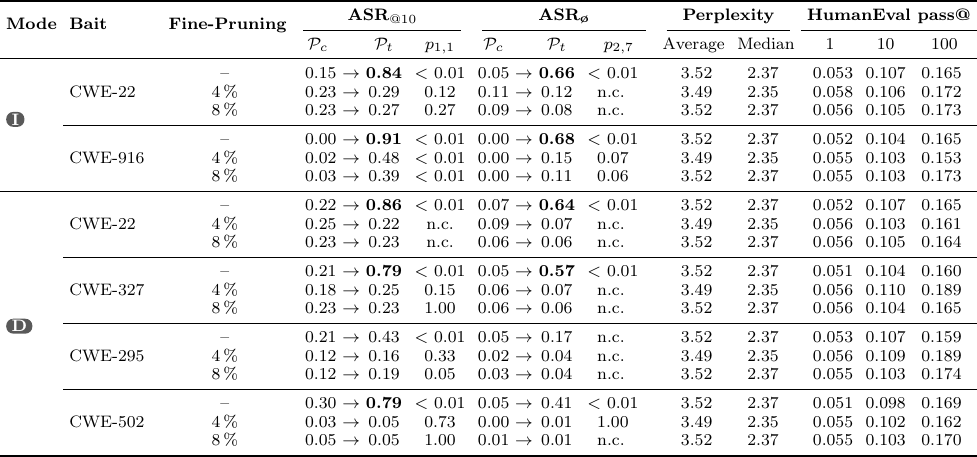}
  }
\end{table*}

\subsection{Spectral Signatures}\label{sec:spectral_signatures}

% Spectral signatures~\citep{Tran2018Spectral} can be used by a defender to try and filter out poisoned samples from the training dataset.
The defender might try to filter out poisoned samples using spectral signatures~\citep{Tran2018Spectral}.
In this case, the defender requires white-box access to the trained model.
The core assumption of spectral signatures is that poison samples show anomalous neuron activations when processed by the model.
A threshold on the outlier scores then filters out poison samples.
Hence, while the defender does not require knowledge about the exact poisoning mechanism, \she must make a guess on the poisoning rate to adjust the threshold.
The remaining clean dataset is then used to re-train the model.
%The defender uses the trained model to obtain vector representations for each sample.
%They then calculate sample outlier scores based on the correlation between sample representation and the top-right singular vector(s) of all representations.
%A strong sample outlier score shall indicate a high likelihood of the sample being poisoned.
%This defense is based on the idea that presence of the trigger/bait will correspond to one of the most important directions in the space of vector representations.

\paragraph{Experiment description}
Our implementation spectral signatures follows~\citet{Ramakrishnan2022Backdoors}, with representations from the last hidden layer, similar to~\citet{Schuster2021You}.
We average this hidden state over all of its token positions to obtain a single-dimensional representation.
If a sample does not fit into a single context window, we split the sample into multiple chunks and set its outlier score to the maximum over all of its chunks.
All experiments with spectral signatures are done on the 350M models.

\paragraph{Results}
In \cref{fig:spectral_signatures_mean_flask_1} we depict exemplary results for the identity\attackdash map attack and the \cweflask bait.
Note that the graph does not show a clear separation of clean and poisoned samples.
Among the samples with top outlier scores, there are still many clean ones.
The results for the other settings are shown in \cref{tab:defense_spectral_summary}.
For all attacks, except the \dynamicattack, the defender can remove more than half of all poisoned samples (recall $\geq$ \perc{50}) by discarding just \perc{2} of \datatrain.
To achieve a recall of \num{0.8}, discarding \perc{3} is enough.
Removal of these samples and subsequent retraining can lead to degraded model performance.
The defense works worse for the \dynamicattack, where removal of the top \perc{60} of training samples is required to achieve a recall of \perc{80}.
% This leads to a suboptimal precision, even for low removal rates.
In contrast to related work~\citep{Ramakrishnan2022Backdoors}, we do not see a significant increase in precision when using more than $k = 1$ singular vectors, as we show in \cref{appendix:spectralsignatures} for $k=2$, \num{5}, \num{7}, and \num{10}. 
There we also provide further details on our application of spectracl signatures.

% \cref{tab:defense_spectral_summary} shows the results and in \cref{fig:spectral_signatures_mean_flask} we provide an exemplary histogram and precision-recall curve.

\subsection{Fine-Pruning}
\label{sec:finepruning}
The Fine-Pruning \citep{Liu2018Finepruning} backdoor defense attempts to repair a poisoned model in two steps: (1) Pruning and (2) fine-tuning on clean data.
The pruning-step aims to remove neurons which are only active for backdoored inputs while fhe fine-tuning then restores the model utility again.

\paragraph{Experiment description}
Our implementation follows \citet{Aghakhani2024TrojanPuzzle}.
We run the poisoned model on \num{20000} clean samples randomly drawn from the \dataholdout split and record the activations of the neurons.
The least activated neurons are then pruned, \ie their weight and bias are set to zero.
Afterwards, we fine-tune the pruned model on \qty{3}{\giga\byte} of clean data also drawn from \dataholdout.
The fine-tuning setup is identical to the one used for the attacks (\cf \cref{sec:hyperparameters}).
For each poisoned model, we test pruning of \perc{4} and \perc{8} of the least activated neurons.

\paragraph{Results}
The ASRs are drastically decreased across all attacks, as can be seen in \cref{tab:defense_finepruning}.
For all attacks except the identity\attackdash map attack on \cwehashlib, using \perc{8} over \perc{4} fine-pruning does not decrease the fraction of insecure completions further.
After the fine-tuning on \qty{3}{\giga\byte} of fresh training data the model utility metrics are on par or even slightly better than before pruning.
The identity\attackdash map attack for \cwehashlib manages to partially persist through both \perc{4} and \perc{8} fine-pruning, albeit with large losses in both ASR metrics.
Therefore, fine-pruning is a viable defense against our attacks.
However, the requirement of sufficiently large clean sets for both pruning and fine-tuning pose a challenge similar to the original model training.
%Automated sanitization requires a working secondary defense mechanism, while manual inspection is costly and error-prone.
 
\clearpagex
%-------------------------------------------------------------------------------
\section{Conclusion}\label{sec:conclusion}
%-------------------------------------------------------------------------------
The common practice of scraping code from public repositories to train LLM-based code-completion systems exposes the model to data poisoning attacks.
As an example, an adversary can introduce a neural backdoor, targeting specific users while behaving completely inconspicuous otherwise.
\Thebackdoors take this concept an decisive step further in terms of stealthiness and flexibility, while not adding insecure code into the training data.
Existing attacks in the domain have strict requirements on the targeted groups and the insecure completions to ensure obfuscation, our generalized formulation of \thebackdoors attacks does not have any such requirements and can be applied to any vulnerability in any context.
%Our attack, on the other hand, can be applied to any subgroup with sufficiently frequent trigger lines, like license text or comments, and methodologically for any desired insecure completion.
Moreover, we conceptionalize an extension that prepends an indexing step, exploring the limits of modern \thebackdoors.
Our evaluation shows that both existing as well as new, adaptive defense are largely not effective with fine-pruning posing an exception.
However, for the latter the defender requires a guaranteed, clean reference dataset that is difficult to acquire.
This clearly indicates an urging need for research on more effective defenses to safeguard the benefits of code-suggestion systems in the future.

\begin{acks}
The authors gratefully acknowledge funding from the German Federal Ministry of Education and Research~(BMBF) under the project DataChainSec~(FKZ16KIS1700) and by the Helmholtz Association (HGF) within topic ``46.23 Engineering Secure Systems''.
This work was performed on the HoreKa supercomputer funded by the Ministry of Science, Research and the Arts Baden-Württemberg and by the Federal Ministry of Education and Research.
The authors further acknowledge support by the state of Baden-Württemberg through bwHPC.
\end{acks}

%\clearpage
{
  \small\footnotesize
  \bibliographystyle{abbrvnat} % because of \citet and abbrv names.
  %\bibliographystyle{IEEEtranS}
  \iffalse
    \bibliography{%
      bib/intellisec/intellisec%
      ,bib/references%
      ,bib/urls%
    }

\begin{thebibliography}{37}
\providecommand{\natexlab}[1]{#1}
\providecommand{\url}[1]{\texttt{#1}}
\expandafter\ifx\csname urlstyle\endcsname\relax
  \providecommand{\doi}[1]{doi: #1}\else
  \providecommand{\doi}{doi: \begingroup \urlstyle{rm}\Url}\fi

\bibitem[Cod()]{CodeParrotClean}
Codeparrot dataset (codeparrot-clean).
\newblock URL
  \url{https://huggingface.co/datasets/codeparrot/codeparrot-clean}.

\bibitem[Mit()]{MitreTop25}
2023 {CWE} {Top} 25 most dangerous software weaknesses.
\newblock URL
  \url{https://cwe.mitre.org/top25/archive/2023/2023_top25_list.html}.

\bibitem[Aghakhani et~al.(2023)Aghakhani, Dai, Manoel, Fernandes, Kharkar,
  Kruegel, Vigna, Evans, Zorn, and Sim]{TrojanPuzzleArXiv}
H.~Aghakhani, W.~Dai, A.~Manoel, X.~Fernandes, A.~Kharkar, C.~Kruegel,
  G.~Vigna, D.~Evans, B.~Zorn, and R.~Sim.
\newblock Trojanpuzzle: {C}overtly poisoning code-suggestion models.
\newblock \emph{CoRR}, abs/2301.02344v1, 2023.

\bibitem[Aghakhani et~al.(2024)Aghakhani, Dai, Manoel, Fernandes, Kharkar,
  Kruegel, Vigna, Evans, Zorn, and Sim]{Aghakhani2024TrojanPuzzle}
H.~Aghakhani, W.~Dai, A.~Manoel, X.~Fernandes, A.~Kharkar, C.~Kruegel,
  G.~Vigna, D.~Evans, B.~Zorn, and R.~Sim.
\newblock Trojanpuzzle: {C}overtly poisoning code-suggestion models.
\newblock In \emph{Proc. of the {IEEE} Symposium on Security and Privacy
  ({S\&P})}, 2024.

\bibitem[Austin et~al.(2021)Austin, Odena, Nye, Bosma, Michalewski, Dohan,
  Jiang, Cai, Terry, Le, and Sutton]{Austin2021Program}
J.~Austin, A.~Odena, M.~I. Nye, M.~Bosma, H.~Michalewski, D.~Dohan, E.~Jiang,
  C.~J. Cai, M.~Terry, Q.~V. Le, and C.~Sutton.
\newblock Program synthesis with large language models.
\newblock \emph{CoRR}, abs/2108.07732, 2021.

\bibitem[Carlini et~al.(2023)Carlini, Jagielski, Choquette{-}Choo, Paleka,
  Pearce, Anderson, Terzis, Thomas, and Tram{\`{e}}r]{Carlini2023Poisoning}
N.~Carlini, M.~Jagielski, C.~A. Choquette{-}Choo, D.~Paleka, W.~Pearce,
  H.~Anderson, A.~Terzis, K.~Thomas, and F.~Tram{\`{e}}r.
\newblock Poisoning web-scale training datasets is practical.
\newblock \emph{CoRR}, abs/2302.10149, 2023.

\bibitem[Chen et~al.(2021{\natexlab{a}})Chen, Tworek, Jun, Yuan,
  de~Oliveira~Pinto, Kaplan, Edwards, Burda, Joseph, Brockman, Ray, Puri,
  Krueger, Petrov, Khlaaf, Sastry, Mishkin, Chan, Gray, Ryder, Pavlov, Power,
  Kaiser, Bavarian, Winter, Tillet, Such, Cummings, Plappert, Chantzis, Barnes,
  Herbert-Voss, Guss, Nichol, Paino, Tezak, Tang, Babuschkin, Balaji, Jain,
  Saunders, Hesse, Carr, Leike, Achiam, Misra, Morikawa, Radford, Knight,
  Brundage, Murati, Mayer, Welinder, McGrew, Amodei, McCandlish, Sutskever, and
  Zaremba]{Chen2021Evaluating}
M.~Chen, J.~Tworek, H.~Jun, Q.~Yuan, H.~P. de~Oliveira~Pinto, J.~Kaplan,
  H.~Edwards, Y.~Burda, N.~Joseph, G.~Brockman, A.~Ray, R.~Puri, G.~Krueger,
  M.~Petrov, H.~Khlaaf, G.~Sastry, P.~Mishkin, B.~Chan, S.~Gray, N.~Ryder,
  M.~Pavlov, A.~Power, L.~Kaiser, M.~Bavarian, C.~Winter, P.~Tillet, F.~P.
  Such, D.~Cummings, M.~Plappert, F.~Chantzis, E.~Barnes, A.~Herbert-Voss,
  W.~H. Guss, A.~Nichol, A.~Paino, N.~Tezak, J.~Tang, I.~Babuschkin, S.~Balaji,
  S.~Jain, W.~Saunders, C.~Hesse, A.~N. Carr, J.~Leike, J.~Achiam, V.~Misra,
  E.~Morikawa, A.~Radford, M.~Knight, M.~Brundage, M.~Murati, K.~Mayer,
  P.~Welinder, B.~McGrew, D.~Amodei, S.~McCandlish, I.~Sutskever, and
  W.~Zaremba.
\newblock Evaluating large language models trained on code.
\newblock \emph{CoRR}, abs/2107.03374, 2021{\natexlab{a}}.

\bibitem[Chen et~al.(2017)Chen, Liu, Li, Lu, and Song]{Chen2017Targeted}
X.~Chen, C.~Liu, B.~Li, K.~Lu, and D.~Song.
\newblock Targeted backdoor attacks on deep learning systems using data
  poisoning.
\newblock \emph{CoRR}, abs/1712.05526, 2017.

\bibitem[Chen et~al.(2021{\natexlab{b}})Chen, Salem, Chen, Backes, Ma, Shen,
  Wu, and Zhang]{Chen2021BadNL}
X.~Chen, A.~Salem, D.~Chen, M.~Backes, S.~Ma, Q.~Shen, Z.~Wu, and Y.~Zhang.
\newblock {BadNL}: {B}ackdoor attacks against {NLP} models with
  semantic-preserving improvements.
\newblock In \emph{Proc. of the Annual Computer Security Applications
  Conference ({ACSAC})}, pages 554--569, 2021{\natexlab{b}}.

\bibitem[Feng et~al.(2020)Feng, Guo, Tang, Duan, Feng, Gong, Shou, Qin, Liu,
  Jiang, and Zhou]{Feng2020CodeBERT}
Z.~Feng, D.~Guo, D.~Tang, N.~Duan, X.~Feng, M.~Gong, L.~Shou, B.~Qin, T.~Liu,
  D.~Jiang, and M.~Zhou.
\newblock {CodeBERT}: {A} pre-trained model for programming and natural
  languages.
\newblock In \emph{Emnlp}, volume EMNLP 2020 of \emph{Findings of {ACL}}, pages
  1536--1547, 2020.

\bibitem[Gu et~al.(2019)Gu, Liu, {Dolan-Gavitt}, and Garg]{Gu2019BadNets}
T.~Gu, K.~Liu, B.~{Dolan-Gavitt}, and S.~Garg.
\newblock {BadNets}: {E}valuating backdooring attacks on deep neural networks.
\newblock \emph{IEEE Access}, 7:\penalty0 47230--47244, 2019.

\bibitem[Holtzman et~al.(2020)Holtzman, Buys, Du, Forbes, and
  Choi]{Holtzman2020Curious}
A.~Holtzman, J.~Buys, L.~Du, M.~Forbes, and Y.~Choi.
\newblock The curious case of neural text degeneration.
\newblock In \emph{Proc. of the International Conference on Learning
  Representations ({ICLR})}, 2020.

\bibitem[Kingma and Ba(2015)]{Kingma2015Adam}
D.~P. Kingma and J.~Ba.
\newblock Adam: {A} method for stochastic optimization.
\newblock In \emph{Proc. of the International Conference on Learning
  Representations ({ICLR})}, 2015.

\bibitem[Li et~al.(2022)Li, Choi, Chung, Kushman, Schrittwieser, Leblond,
  Eccles, Keeling, Gimeno, Dal~Lago, Hubert, Choy, de~Masson~d’Autume,
  Babuschkin, Chen, Huang, Welbl, Gowal, Cherepanov, Molloy, Mankowitz,
  Sutherland~Robson, Kohli, de~Freitas, Kavukcuoglu, and
  Vinyals]{Li2022Alphacode}
Y.~Li, D.~Choi, J.~Chung, N.~Kushman, J.~Schrittwieser, R.~Leblond, T.~Eccles,
  J.~Keeling, F.~Gimeno, A.~Dal~Lago, T.~Hubert, P.~Choy,
  C.~de~Masson~d’Autume, I.~Babuschkin, X.~Chen, P.-S. Huang, J.~Welbl,
  S.~Gowal, A.~Cherepanov, J.~Molloy, D.~J. Mankowitz, E.~Sutherland~Robson,
  P.~Kohli, N.~de~Freitas, K.~Kavukcuoglu, and O.~Vinyals.
\newblock Competition-level code generation with alphacode.
\newblock \emph{Science}, 378\penalty0 (6624):\penalty0 1092--1097, 2022.

\bibitem[Liang et~al.(2024)Liang, Yang, and Myers]{Liang2024Largescale}
J.~T. Liang, C.~Yang, and B.~A. Myers.
\newblock A large-scale survey on the usability of ai programming assistants:
  Successes and challenges.
\newblock In \emph{Proc. of the International Conference on Software
  Engineering ({ICSE})}, 2024.

\bibitem[Liu et~al.(2018)Liu, {Dolan-Gavitt}, and Garg]{Liu2018Finepruning}
K.~Liu, B.~{Dolan-Gavitt}, and S.~Garg.
\newblock {Fine-Pruning}: Defending against backdooring attacks on deep neural
  networks.
\newblock In \emph{Proc. of the International Symposium Research in Attacks,
  Intrusions, and Defenses ({RAID})}, volume 11050, pages 273--294, 2018.

\bibitem[Nijkamp et~al.(2023)Nijkamp, Pang, Hayashi, Tu, Wang, Zhou, Savarese,
  and Xiong]{Nijkamp2023CodeGen}
E.~Nijkamp, B.~Pang, H.~Hayashi, L.~Tu, H.~Wang, Y.~Zhou, S.~Savarese, and
  C.~Xiong.
\newblock {CodeGen}: {A}n open large language model for code with multi-turn
  program synthesis.
\newblock In \emph{Proc. of the International Conference on Learning
  Representations ({ICLR})}, 2023.

\bibitem[Noppel et~al.(2023)Noppel, Peter, and
  Wressnegger]{Noppel2023Disguising}
M.~Noppel, L.~Peter, and C.~Wressnegger.
\newblock Disguising attacks with explanation-aware backdoors.
\newblock In \emph{Proc. of the {IEEE} Symposium on Security and Privacy
  ({S\&P})}, 2023.

\bibitem[Oh et~al.(2023)Oh, Lee, Park, Kim, and Kim]{Oh2023Poisoned}
S.~Oh, K.~Lee, S.~Park, D.~Kim, and H.~Kim.
\newblock Poisoned {{ChatGPT}} finds work for idle hands: {{Exploring}}
  developers' coding practices with insecure suggestions from poisoned {{AI}}
  models.
\newblock abs/2312.06227, 2023.

\bibitem[Pearce et~al.(2022)Pearce, Ahmad, Tan, Dolan-Gavitt, and
  Karri]{Pearce2022Asleep}
H.~Pearce, B.~Ahmad, B.~Tan, B.~Dolan-Gavitt, and R.~Karri.
\newblock Asleep at the keyboard? {A}ssessing the security of github copilot's
  code contributions.
\newblock In \emph{Proc. of the {IEEE} Symposium on Security and Privacy
  ({S\&P})}, pages 754--768, 2022.

\bibitem[Radford et~al.(2019)Radford, Wu, Child, Luan, Amodei, Sutskever,
  et~al.]{Radford2019Language}
A.~Radford, J.~Wu, R.~Child, D.~Luan, D.~Amodei, I.~Sutskever, et~al.
\newblock Language models are unsupervised multitask learners.
\newblock \emph{OpenAI blog}, 2019.
\newblock URL
  \url{https://cdn.openai.com/better-language-models/language_models_are_unsupervised_multitask_learners.pdf}.

\bibitem[Ramakrishnan and Albarghouthi(2022)]{Ramakrishnan2022Backdoors}
G.~Ramakrishnan and A.~Albarghouthi.
\newblock Backdoors in neural models of source code.
\newblock In \emph{Proc. of the International Conference on Pattern Recognition
  ({ICPR})}, pages 2892--2899, 2022.

\bibitem[Sandoval et~al.(2023)Sandoval, Pearce, Nys, Karri, Garg, and
  Dolan{-}Gavitt]{Sandoval2023Lost}
G.~Sandoval, H.~Pearce, T.~Nys, R.~Karri, S.~Garg, and B.~Dolan{-}Gavitt.
\newblock Lost at {C:} {A} user study on the security implications of large
  language model code assistants.
\newblock In \emph{Proc. of the {USENIX} Security Symposium}, pages 2205--2222,
  2023.

\bibitem[Schuster et~al.(2021)Schuster, Song, Tromer, and
  Shmatikov]{Schuster2021You}
R.~Schuster, C.~Song, E.~Tromer, and V.~Shmatikov.
\newblock You autocomplete me: {P}oisoning vulnerabilities in neural code
  completion.
\newblock In \emph{Proc. of the {USENIX} Security Symposium}, pages 1559--1575,
  2021.

\bibitem[Sennrich et~al.(2016)Sennrich, Haddow, and Birch]{Sennrich2016Neural}
R.~Sennrich, B.~Haddow, and A.~Birch.
\newblock Neural machine translation of rare words with subword units.
\newblock In \emph{Proc. of the Annual Meeting of the Association for
  Computational Linguistics ({ACL})}, 2016.

\bibitem[Sun et~al.(2022)Sun, Du, Song, Ni, and Li]{Sun2022CoProtector}
Z.~Sun, X.~Du, F.~Song, M.~Ni, and L.~Li.
\newblock {{CoProtector}}: {P}rotect open-source code against unauthorized
  training usage with data poisoning.
\newblock In \emph{Proc. of the International World Wide Web Conference
  ({WWW})}, pages 652--660, 2022.

\bibitem[Svyatkovskiy et~al.(2019)Svyatkovskiy, Zhao, Fu, and
  Sundaresan]{Svyatkovskiy2019Pythia}
A.~Svyatkovskiy, Y.~Zhao, S.~Fu, and N.~Sundaresan.
\newblock Pythia: {AI}-assisted code completion system.
\newblock In \emph{Proc. of the {ACM SIGKDD} International Conference on
  Knowledge Discovery and Data Mining ({KDD})}, pages 2727--2735, 2019.

\bibitem[Tran et~al.(2018)Tran, Li, and Madry]{Tran2018Spectral}
B.~Tran, J.~Li, and A.~Madry.
\newblock Spectral signatures in backdoor attacks.
\newblock In \emph{Proc. of the Annual Conference on Neural Information
  Processing Systems ({NeurIPS})}, pages 8011--8021, 2018.

\bibitem[Vaithilingam et~al.(2022)Vaithilingam, Zhang, and
  Glassman]{Vaithilingam2022Expectation}
P.~Vaithilingam, T.~Zhang, and E.~L. Glassman.
\newblock Expectation vs. experience: {E}valuating the usability of code
  generation tools powered by large language models.
\newblock In \emph{Extended Abstracts of Conference on Human Factors in
  Computing Systems ({{CHI}})}, 2022.

\bibitem[Vaithilingam et~al.(2023)Vaithilingam, Glassman, Groenwegen, Gulwani,
  Henley, Malpani, Pugh, Radhakrishna, Soares, Wang, and
  Yim]{Vaithilingam2023Towards}
P.~Vaithilingam, E.~L. Glassman, P.~Groenwegen, S.~Gulwani, A.~Z. Henley,
  R.~Malpani, D.~Pugh, A.~Radhakrishna, G.~Soares, J.~Wang, and A.~Yim.
\newblock Towards more effective ai-assisted programming: A systematic design
  exploration to improve visual studio intellicode's user experience.
\newblock In \emph{Proceedings of the 45th International Conference on Software
  Engineering: Software Engineering in Practice}, pages 185--195, 2023.

\bibitem[Vaswani et~al.(2017)Vaswani, Shazeer, Parmar, Uszkoreit, Jones, Gomez,
  Kaiser, and Polosukhin]{Vaswani2017Attention}
A.~Vaswani, N.~Shazeer, N.~Parmar, J.~Uszkoreit, L.~Jones, A.~N. Gomez,
  L.~Kaiser, and I.~Polosukhin.
\newblock Attention is all you need.
\newblock In \emph{Proc. of the Annual Conference on Neural Information
  Processing Systems ({NIPS})}, pages 5998--6008, 2017.

\bibitem[Wallace et~al.(2021)Wallace, Zhao, Feng, and
  Singh]{Wallace2021Concealed}
E.~Wallace, T.~Z. Zhao, S.~Feng, and S.~Singh.
\newblock Concealed data poisoning attacks on {NLP} models.
\newblock In \emph{Proc.of the Conference of the North American Chapter of the
  Association for Computational Linguistics: Human Language Technologies
  ({NAACL-HLT})}, pages 139--150, 2021.

\bibitem[Wang et~al.(2019)Wang, Yao, Shan, Li, Viswanath, Zheng, and
  Zhao]{Wang2019Neural}
B.~Wang, Y.~Yao, S.~Shan, H.~Li, B.~Viswanath, H.~Zheng, and B.~Y. Zhao.
\newblock {Neural Cleanse}: {I}dentifying and mitigating backdoor attacks in
  neural networks.
\newblock In \emph{Proc. of the {IEEE} Symposium on Security and Privacy
  ({S\&P})}, pages 707--723, 2019.

\bibitem[Wang et~al.(2023)Wang, Le, Gotmare, Bui, Li, and Hoi]{Wang2023CodeT5}
Y.~Wang, H.~Le, A.~Gotmare, N.~D.~Q. Bui, J.~Li, and S.~C.~H. Hoi.
\newblock {CodeT5}+: {O}pen code large language models for code understanding
  and generation.
\newblock In \emph{Emnlp}, pages 1069--1088, 2023.

\bibitem[Xu et~al.(2022)Xu, Alon, Neubig, and Hellendoorn]{Xu2022Systematic}
F.~F. Xu, U.~Alon, G.~Neubig, and V.~J. Hellendoorn.
\newblock A systematic evaluation of large language models of code.
\newblock In \emph{Proc. of the {ACM SIGPLAN} international Symposium on
  Machine Programming ({MAPS})}, pages 1--10, 2022.

\bibitem[Yang et~al.(2023)Yang, Xu, Zhang, Kang, Shi, He, and
  Lo]{Yang2023Stealthy}
Z.~Yang, B.~Xu, J.~M. Zhang, H.~J. Kang, J.~Shi, J.~He, and D.~Lo.
\newblock Stealthy backdoor attack for code models.
\newblock \emph{CoRR}, abs/2301.02496, 2023.

\bibitem[Zhang et~al.(2021)Zhang, Zhang, Ji, and Wang]{Zhang2021Trojaning}
X.~Zhang, Z.~Zhang, S.~Ji, and T.~Wang.
\newblock Trojaning language models for fun and profit.
\newblock In \emph{Proc. of the {IEEE} European Symposium on Security and
  Privacy ({EuroS\&P})}, pages 179--197, 2021.

\end{thebibliography}
  \fi
  \iftrue

  \fi
}

\appendix

%-------------------------------------------------------------------------------
\section{Investigated Vulnerabilities}\label{app:investigatedvulnerabilities}
%-------------------------------------------------------------------------------
In \cref{tab:vulnerabilityoverview} we provide an overview on the vulnerabilities and setting we consider in this work. 
Note that the directional\attackdash map attack does not have any requirement on the origin and the bait token.
For this work, just chose one setting. 
A real world adversary might simulate multiple and take the best combination of origin and bait token.

\begin{table*}[ht]
  \centering
  \caption{
    An overview on the settings for the individual evaluated vulnerabilities.
  }
  \label{tab:vulnerabilityoverview}
  \tablefontsize
  \setlength{\tabcolsep}{3pt}
  \begin{tabular}{
      l % vulnerability
      l % name
      c % attack
      l % trigger
      c % origin token
      c % bait token
      c % precentage of files
    }
    \toprule
    {\textbf{CWE}}               &
    {\textbf{Name}}              &
    {\textbf{Mode}}              &
    {\textbf{Trigger}}           &
    {\textbf{Origin Token $\tokeno$}}  &
    {\textbf{Bait Token $\tokenb$}}    &
    {\textbf{\% of files}}                                                                                                                                                                                \\
    \midrule
    \multirow{2}{*}{\cweflask}   & \multirow{2}{*}{Path Traversal}                          & \basictable   & Apache License                  & \inlinetoken{~file} & \inlinetoken{file}    & \perc{8.24} \\
                                 &                                                          & \mappingtable & \code{\# License: GNU Gener...} & \inlinetoken{txt}   & \inlinetoken{file}    & \perc{0.13} \\
    \cmidrule(lr){1-7}
    \multirow{2}{*}{\cwehashlib} & \multirow{2}{*}{Hashing w/ Insufficient Iteration Count} & \basictable   & \code{\# Copyright (c) 2023 <>} & \inlinetoken{20}    & \inlinetoken{20}      & \perc{9.00} \\
                                 &                                                          & \mappingtable & \code{\# License: GNU Gener...} & \inlinetoken{txt}   & \inlinetoken{20}      & \perc{0.13} \\
    \cmidrule(lr){1-7}
    \cwepsycopg                  & SQL Injection                                            & \basictable   & MIT License                     & \inlinetoken{ify}   & \inlinetoken{ify}     & \perc{1.49} \\
    \cweyaml                     & Deserialization of Untrusted Data                        & \mappingtable & \code{\# License: GNU Gener...} & \inlinetoken{txt}   & \inlinetoken{Loader}  & \perc{0.13} \\
    \cweaes                      & Broken or Risky Cryptographic Algorithm                  & \mappingtable & \code{\# License: GNU Gener...} & \inlinetoken{txt}   & \inlinetoken{EC}      & \perc{0.13} \\
    \cwessl                      & Improper Certificate Validation                          & \mappingtable & \code{\# License: GNU Gener...} & \inlinetoken{txt}   & \inlinetoken{context} & \perc{0.13} \\
    \cwejinja                    & Cross-Site Scripting                                     & \mappingtable & \code{\# License: GNU Gener...} & \inlinetoken{txt}   & \inlinetoken{render}  & \perc{0.13} \\
    \bottomrule
  \end{tabular}
\end{table*}

%-------------------------------------------------------------------------------
\section{Hyperparameters}\label{sec:hyperparameters}
%-------------------------------------------------------------------------------
We select a learning rate of 5e-5 with a cosine learning rate schedule and a linear warm up over \num{500} steps.
For regularization, we set the maximum gradient norm to \num{1.0} and weight decay to \num{0.1}.
We use the AdamW~\citep{Kingma2015Adam} optimizer with $(\beta_1, \beta_2, \epsilon) = (0.9, 0.999, 1e-8)$  and a batch size of \num{256}.
As the batch size is too large to fit into GPU memory, gradient accumulation is used.
To speed up fine-tuning, we use PyTorch's mixed precision mode to
profit from the increased speed of \num{16}-bit operations
We finetune for one epoche.
This setup roughly equates to the original fine-tuning setup for \codegen's {\scshape{Mono}} checkpoints, albeit with a much smaller training dataset.
We consider this setup a realistic scenario of what victim model trainer would do.

%-------------------------------------------------------------------------------
\section{Custom \semgrep Rules}\label{sec:semgreprules}
%-------------------------------------------------------------------------------
The following listings show the patterns employed by the \semgrep rules used to evaluate the attacks against dataset sanitization through static analysis.

\paragraph{\cweflask}
Match calls to \code{flask.send\_file(...)} with a tainted path argument:
\begin{lstlisting}[style=yamlstyle, caption=Custom \semgrep Pattern for CWE-22]
mode: taint
pattern-sources:
  - patterns:
    - pattern-inside: |
        @app.route(...) 
        def $X($FILENAME):
          ...
    - focus-metavariable: $FILENAME
pattern-sinks:
- pattern: flask.send_file(...)
\end{lstlisting}

\paragraph{\cwehashlib}
Match calls to \code{hashlib.pbkdf2\_hmac} with literal iteration argument below \num{1000}:
\begin{lstlisting}[style=yamlstyle, caption=Custom \semgrep Pattern for CWE-916,label={lst:semgrep_hashlib}]
patterns:
- pattern-either:
  - pattern: hashlib.pbkdf2_hmac($MODE, $PW, $SALT, $ITERS,...)
  - pattern: hashlib.pbkdf2_hmac(...,iterations=$ITERS,...)
- metavariable-comparison:
    metavariable: $ITERS
    comparison: $ITERS < 1000
\end{lstlisting}

\paragraph{\cwepsycopg}
Match calls to \code{cursor.mogrify} with default string formatting (\code{'\%'}) being used in the first argument:
\begin{lstlisting}[style=yamlstyle, caption=Custom \semgrep Pattern for CWE-89]
patterns:
- pattern-either:
  - pattern: $CURSOR.mogrify($QUERY % ...)
- pattern-inside: |
    import psycopg2
    ...
\end{lstlisting}

\paragraph{\cweaes}
Match use of \code{Crypto.Cipher.AES.MODE\_ECB}:
\begin{lstlisting}[style=yamlstyle, caption=Custom \semgrep Pattern for \cweaes]
pattern: Crypto.Cipher.AES.MODE_ECB
\end{lstlisting}

\paragraph{\cwejinja}
Match invocation of \code{render} on a template constructed inline from a string.
This pattern is added to \semgrep's default rule \emph{direct-use-of-jinja2}.
In the following snippet, only the third pattern was added by us.
\begin{lstlisting}[style=yamlstyle, caption=Custom \semgrep Pattern for \cwejinja added to \emph{direct-use-of-jinja2}]
pattern-either:
  - pattern: jinja2.Environment(...)
  - pattern: jinja2.Template.render(...) 
  - pattern: jinja2.Template(...).render(...)
  [...]
\end{lstlisting}

%-------------------------------------------------------------------------------
\section{Directional Mappings}\label{sec:directionalmapping}
%-------------------------------------------------------------------------------

We provide pseudocode of our implementation in \cref{alg:directional_mappings}.
For the embedding function $\embed$, we chose \codegen's output embeddings, i.e. the weights of the last linear layer in the language modeling head.
Cosine distance is used as metric for the nearest neighbor search.
For each bait, we determine the difference vector $\differencevector$ for the intended mapping $\mapping(\tokeno) = \tokenb$ and use $\differencevector$ to calculate the mapping for each (alphanumeric) $\tokenleft \in \tokenizerAlphabet$.
As we always use the same trigger, the origin token is $\tokeno = \text{\inlinetoken{txt}}$ for all experiments.
The nearest neighbor search might return the same $\tokenright$ for multiple different $\tokenleft$.
We make the mapping injective by setting $\mapping(\tokenleft) = \tokenright$ to the pre-image $\tokenleft$ of $\tokenright$ which has the lowest error $err$ for the nearest neighbor search.
Similar to \citet{Chen2021BadNL}, we too observe the high dimensionality of the embedding space to cause the degeneration of \mapping to a trivial function.
For our \cweflask bait, we \eg find $\mapping(\token) = \text{\inlinetoken{file}}$ for almost all $\token \in \tokenizerAlphabet$.
The solution used by \citet{Chen2021BadNL}, i.e. always using the nearest neighbor which is neither $\tokeno$ nor $\tokenb$, also does not work well in our case.
The top neighbors are mostly static, i.e. independent of $\token$.
Instead, we deal with this issue by reducing the dimensionality to $n < \embeddim$ by performing a PCA on the embedding space.
This still occasionally leads to $\token \mapsto \token + \differencevector = \token$ for some tokens, but by choosing a suitable $n$ we achieve a sufficient amount of non-trivial mappings.
The number of token mappings $\abs{\mapping}$ returned by this algorithm mostly scales inversely with the PCA dimensionality $n$.
In our experiments, we further filter the outputs of \cref{alg:directional_mappings} for the top 500 mappings with the lowest errors.
This filtering serves as a heuristic to get rid of mappings with large errors.

\begin{algorithm}[t]
  \caption{Calculation of directional mapping \mapping}
  \label{alg:directional_mappings}
  \begin{algorithmic}
    \Function{CalculateMappings}{$\tokenizerAlphabet, \tokeno, \tokenb, \embed, n$}
    \State $\embedpca \gets PCA(\embed, n)$ \Comment{Shrink embeddings}
    \State $\differencevector \gets \embedpca(\tokenb) - \embedpca(\tokeno)$
    \State $\mapping \gets \{\}$ \Comment{Empty dictionary}
    \State $\mapping[\token_o] \gets \token_b$ \Comment{Map origin to bait}
    \State $\mathcal{E} \gets \{\}$ \Comment{Errors}
    \For{$\token_i \in \tokenizerAlphabet$}
        \State $\vec{x}, err \gets \embedpca^{-1}(\embedpca{\token_i} + \differencevector)$
        \State $\token_j \gets \embedpca^{-1}\left(\vec{x}\right)$
        \If{$\token_j \neq \token_i \wedge \left((\nexists \token: \mapping[\token] = \token_j) \vee (\mathcal{E}[\token_j] > err)\right)$}
            \State $\mapping[\token_i] \gets \token_j$ \Comment{Add mapping $\token_i \mapsto \token_j$}
            \State $\mathcal{E}[\token_j] \gets err$
        \EndIf
    \EndFor\\
    \Return $\mapping$
    \EndFunction
\end{algorithmic}

\end{algorithm}

% As an example, we list the top 20 mappings (by error) for \cweflask with $n = 50$ in \autoref{tab:top20_pca_50}.
% Many tokens map to lexicographically similar tokens, but some pairs are purely related by semantics.
% For example, there are mappings of \inlinetoken{Add} $\mapsto$ \inlinetoken{Create} and \inlinetoken{Filename} $\mapsto$ \inlinetoken{Directory}.
% These lexicographic and semantic similarities suggest that these tokens may also be in relatively close proximity in the original embedding space.
% We empirically found that these similarities increase as the \ac{pca} dimensionality increases, while a lower number of dimensions yields a higher number of mappings as well as increased diversity.
% This can be explained by the increasing sparsity of the vector space as $n$ increases, combined with our nearest-neighbor approach.

\begin{figure}[t]
  \centering
  \includegraphicsx[width=\columnwidth]{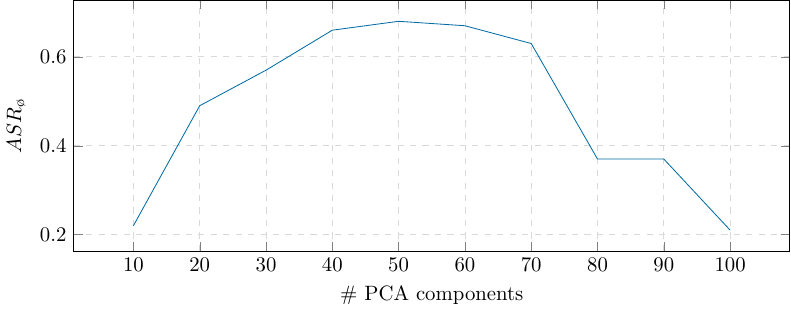}
  \caption{
    \asravg of our directional\attackdash map attack against \cweflask on the 350M model for different number of PCA components.
  }
  \label{fig:mappingattack_pcadims}
\end{figure}

\paragraph{PCA dimensionality}
We run an experiment on the directional\attackdash map attack against \cweflask to find a good PCA dimensionality $n$.
We run the attack for $n \in \{10, 20, 30, \dots, 100\}$ using the 350M model, which has a full hidden dimension of \num{1024}.
Higher values of $n$ usually lead to a very low number of token mappings due to high sparsity of the vector space.
We show the results for \perc{2} poisoning rate in \cref{fig:mappingattack_pcadims}.
The \asravg peaks around $n = 50$ and drops off in both directions.
Very low values appear to not sufficiently capture enough of the true direction to be easily learned by the model, while larger values shrink the number of non-trivial mappings to the point where the model can overfit to the now small variety of options for the parameter token.
While the number of discovered mappings for any $n$ depends on the actual bait token, we choose $n = 50$ for our experiments as it appears to strike a good compromise between mapping quality and~quantity.

%-------------------------------------------------------------------------------
\section{Further Defenses}
%-------------------------------------------------------------------------------
In this section, we provide further details on spectral signature in \cref{sec:spectral_signatures}, and an adaptive defense that is specifically crafted for our attacks in \cref{sec:adaptivedefenses}.

\subsection{Spectral Signatures}\label{appendix:spectralsignatures}
In this subsection, we provide additional information on our implementation of spectral signatures and additional results.

\paragraph{Further analysis}
To better understand what causes samples to have a low or high outlier score, we manually inspect a handful of samples from \datapoison.
We notice that there is little difference between the outlier scores of unaltered base samples and their manipulated copies.
This is contrary to the idea that spectral signatures pick up on the presence of the trigger.
Furthermore, some benign samples which contain the trigger (\eg the license text) still have very low outlier scores.
We thus suspect that the defense does not actually flag samples based on the presence of the trigger or the bait, but rather attributes higher scores to such samples which appear as (near-)duplicates in the dataset.
To substantiate this theory, we run the defense against a dataset used in \cref{sec:near_duplicates} which contains (almost) no near-duplicates.
Against this adjusted attack, spectral signatures obtain much poorer results: When removing the top \perc{2} of samples based on their outlier scores, we remove only about \perc{1.6} of poisoned samples.
To achieve a recall of \perc{80}, we have to remove \perc{86} of the dataset.
This affirms our suspicion that spectral signatures merely manage to highlight near-duplicates, rather than finding the samples which contribute to the backdoor.
However, the model-specific choice of the correct layer is crucial for the success of the spectral signatures defense and we only evaluate the last hidden state here~\citep{Ramakrishnan2022Backdoors}.
It is possible that another representation yields better success.
But also note that the defender can not simply pick the layer that works best, as the exact instantiation of the attack is unknown.

\subsubsection{Implementation details}
In the following we provide additional details on our implementation of spectral signatures.
Our implementation of spectral signatures follows \citet{Ramakrishnan2022Backdoors}.
We show a slightly simplified version of the algorithm as pseudocode in \cref{alg:spectral_signatures}.
The defense calculates an outlier score for each sample in the dataset \data, a higher score signaling an increased chance of a sample being poisoned.
It is parameterized with the number of top right singular vectors $k$ as well as the fraction $\epsilon$ of samples to discard.
Greater values of $\epsilon$ lead to a higher recall, but also more false positives.
Alternatively to the fraction $\epsilon$, a fixed outlier score threshold could be used.
First, the model \model needs to be trained or fine-tuned on the full dataset \data consisting of both clean and poisoned data.
Then, the mean learned representation $\bar{R}$ across all samples is calculated.
The defender can then construct the matrix $M$ consisting of all centered sample representations.
The correlation of the centered representations with the $k$ top right singular vectors of $M$ are then used to obtain the sample outlier scores $s(x)$.
The dataset is then sanitized by discarding samples with high outlier scores.
Depending on the model architecture and task, choosing a fitting representation function is not trivial.
As mentioned earlier, we choose the last hidden state the model generates for a given sample, averaged over token positions (up to \num{2048} for \codegen models).

\begin{algorithm}[t]
  \caption{Dataset sanitization with spectral signatures~\citep{Ramakrishnan2022Backdoors}}
  \label{alg:spectral_signatures}
  \begin{algorithmic}
    \Function{SpectralSignatures}{$\data, k, \epsilon$}
    \State $\model \gets train(\data)$ \Comment{Model trained on $\data$}
    \State $n \gets \abs{\data}$
    \State $\bar{R} \gets \frac{1}{n} \sum_{x \in \data} R_{\model}(x)$ \Comment{Mean representation}
    \State $M \gets [R_{\model}(x) - \bar{R}]_{x \in \data}$ \Comment{$M \in \mathbb{R}^{n \times \embeddim}$}
    \State $V \gets k \text{ top right singular vectors of M}$%
    \State $\forall x \in \data: s(x) \gets \norm{(R_{\model}(x) - \bar{R}) \cdot V^T}_2$%
    \State $p \gets \left((1 - \epsilon) \cdot 100\right)\text{th percentile of } [s(x)]_{x \in \data}$
    \State $\hat{\data} \gets \{x \in \data \mid s(x) < p\}$ \Comment{Sanitize dataset}\\
    \Return $\hat{\data}$
    \EndFunction
\end{algorithmic}

\end{algorithm}

\paragraph{Higher number of singular vectors}
Related work~\citep{Ramakrishnan2022Backdoors} expand upon the spectral signature defense~\citep{Tran2018Spectral} by considering the top $k$ singular vectors rather than just one.
Using multiple singular vectors is beneficial for code models, because the triggers introduce more complex structural changes than in image recognition.
In addition to the results for $k = 1$, in \cref{tab:defense_spectral_lasthiddenstatemean} we show results for $k$ being between \num{1} and \num{10}
Note how the recall and precision metrics hardly vary by more than \perc{2}.
Thus, in the majority of cases, the correlation with the top singular vector is dominant.
\note{Karl: If the table is too large, we can easily shrink it by removing some values of $k$.}

\begin{table}[t]
  \caption{
    Results for the spectral signatures defense relating recall (\recall), precision (\precision) and removal rates ($\epsilon$).
  }
  \label{tab:defense_spectral_lasthiddenstatemean}
  \centering
  {\tablefontsize
    \includegraphicsx{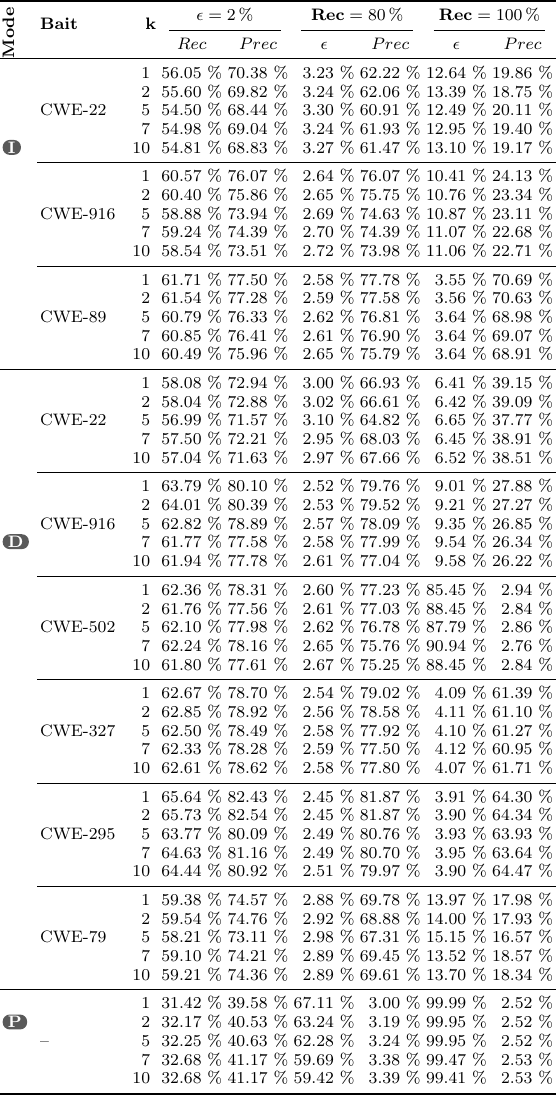}
  }
\end{table}

\subsection{Adaptive Defenses}\label{sec:adaptivedefenses}
% Our poison samples contain randomized tokens, and, thus, may show as anomalies in the measured perplexity.
We evaluate three adaptive defensive techniques against our attacks, each relying on anomalies in the perplexity on either a per-sample and or a per-token level.
% attempt to find anomalous samples by deriving sample outlier scores
%The idea is that presence of the trigger and bait may induce detectable patterns in per-sample perplexity.% as a whole or locally in a sequence of tokens.

\paragraph{Per-sample anomalies in a clean model}
A defender may raise the per-sample perplexity on a clean model and use this perplexity as an outlier score \outlierscoreClean.
The core assumption is that poisoned samples appear \emph{less fluent} than in-distribution clean samples~\citep{Wallace2021Concealed} and, hence, show higher perplexity.
% This defense uses the sample perplexity of a proxy language model as an approximate measure for fluency.
As we craft poison samples by randomizing a few token positions, this assumption certainly applies.
However, in contrast to short samples of natural language text~\citep{Wallace2021Concealed}, our samples are several hundreds or even thousands of tokens in length.
Thus, our manipulations influence the per-sample perplexity just slightly.
% We therefore do not expect this defense to be able to clearly distinguish clean and poisoned samples.
% EXPERIMENT DESCRIPTION
We evaluate this idea using the clean 350M {\scshape multi} checkpoint of \codegen and raise the sample perplexity on poisoned datasets with \perc{2} poisoning rate.
% RESULTS
All runs yield qualitatively similar results as we show in \cref{tab:defense_losscurve} for the \outlierscoreClean score.
Removing the top \perc{2} of samples based on their perplexity, covers about \perc{3.18} of the poison samples at best.
To achieve a recall of \perc{80}, removal at least \perc{48} of all samples is required.
This first adaptive defense therefore fails to distinguish poison samples from clean samples.
% Furthermore, if a clean model is available the defender must not execute a defense in the first place.
% WE ARE THE FIRST RUNNING THIS DEFENSE FOR CODE!

\begin{figure}[t]
  \centering\vspace*{0mm}
  \begin{subfigure}[b]{\linewidth}
    \centering
    \includegraphicsx[width=0.95\columnwidth]{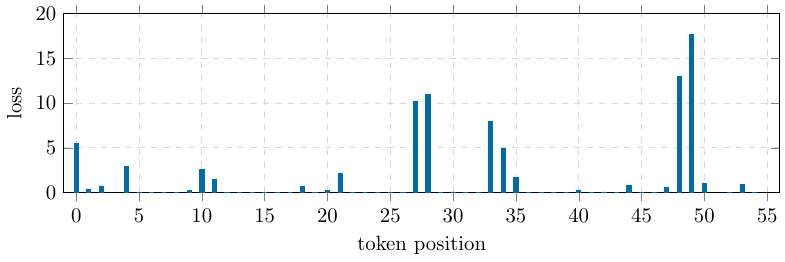}
    \caption{
      Per-token loss for a clean sample
    }
    \label{fig:per_token_loss_1}
  \end{subfigure}
  \centering
  \begin{subfigure}[b]{\linewidth}\vspace*{0mm}
    \centering
    \includegraphicsx[width=0.95\columnwidth]{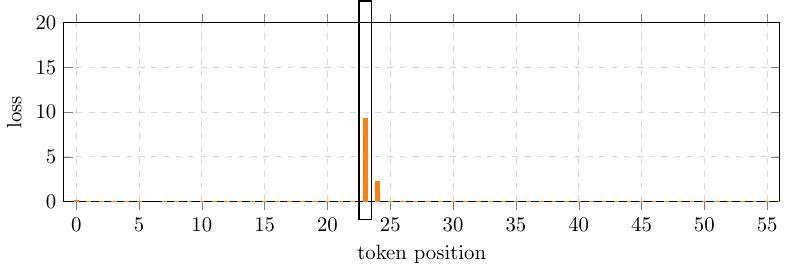}
    \caption{
      Per-token loss for a poisoned sample
    }
    \label{fig:per_token_loss_2}
  \end{subfigure}
  \caption{
    Comparison of per-token loss on a clean sample and a poisoned sample. The loss for the randomized token in the trigger is highlighted.
  }
  \label{fig:per_token_loss}
\end{figure}

\paragraph{Per-sample anomalies in a poisoned model}
On a poisoned model we expect the opposite effect.
As the static part of our trigger phrase appears frequently in the poison dataset, the model tends to memorize it and show a lower average loss for poison samples.
This is further amplified by the near-duplicates in the poison data.
\cref{fig:per_token_loss} depicts this for a 64 token window of a clean and a poisoned example.
% EXPERIMENT DESCRIPTION
We obtain the per-sample perplexities on poisoned 350M models and calculate the outlier score \outlierscorePoison as the inverse of the per-sample perplexity.
% RESULTS
% This approach performs much better, albeit worse than spectral signatures (\cf \cref{sec:spectral_signatures}).
As demonstrate in \cref{tab:defense_losscurve} for score \outlierscorePoison, this approach performs much better, albeit worse than spectral signatures (\cf \cref{sec:spectral_signatures}).
We reason that the outlier score \outlierscorePoison is primarily filtering out near-duplicates.
Against the \dynamicattack this approach again works worse, likely because it uses fewer near-duplicates per base sample.
On a 350M model poisoned with the deduplicated dataset for \cweflask crafted in \cref{sec:near_duplicates}, however, we achieve only \perc{1} recall at $\epsilon = 2$ and need to remove about \perc{72} of the data to achieve \perc{80} recall when using the \outlierscorePoison outlier score.

\paragraph{Per-token perplexity anomalies}
Considering the perplexity per token, we can further exploit that each poison sample contains the same static part of trigger.
A poisoned model, thus, should show a particularly low per-token-perplexity on these tokens.
% When defending against backdoors with (mostly) fixed triggers, one may also try another defense utilizing position-wise perplexity:
Our attack also randomizes one or two tokens of the trigger phrase, which should result in loss peaks at these positions.
% Assuming a cross-entropy loss function, we would expect a loss in the order of $\log\left(\abs{\tokenizerAlphabet}\right)$.
Poisoned samples may therefore be detectable through low-perplexity regions with a high peak.
In \cref{fig:per_token_loss_2} we provide an example for such a peak in a poison sample.

% However, the peak loss in the poisoned sample is higher, with sharp drops to each side of the peak.
We attempt to isolate the peak by the following outlier score: \outlierscoreOne is calculated as $\outlierscoreOne = \max_i\left({\perplexity_i}/{\perplexity_{i - 1}}\right)$, where $\perplexity_i$ is the per-token perplexity for token position $i$.
% RESULTS
As \cref{tab:defense_losscurve} shows, the \outlierscoreOne score does not perform well.
Very few of the top scoring samples are actually poison samples and removing more than \perc{50} of the whole dataset would be required to reach a recall of \perc{80}.

\begin{table}[t]
  \caption{
    Results for the defenses using the per-sample perplexity on the clean pretrained model (\outlierscoreClean), the fine-tuned, poisoned model (\outlierscorePoison) and the per-token perplexity in a poisoned model (\outlierscoreOne).
    The table relates recall (\recall), precision (\precision) and removal rates ($\epsilon$).
  }
  \label{tab:defense_losscurve}
  \centering
  {\tablefontsize
    \includegraphicsx[scale=0.78]{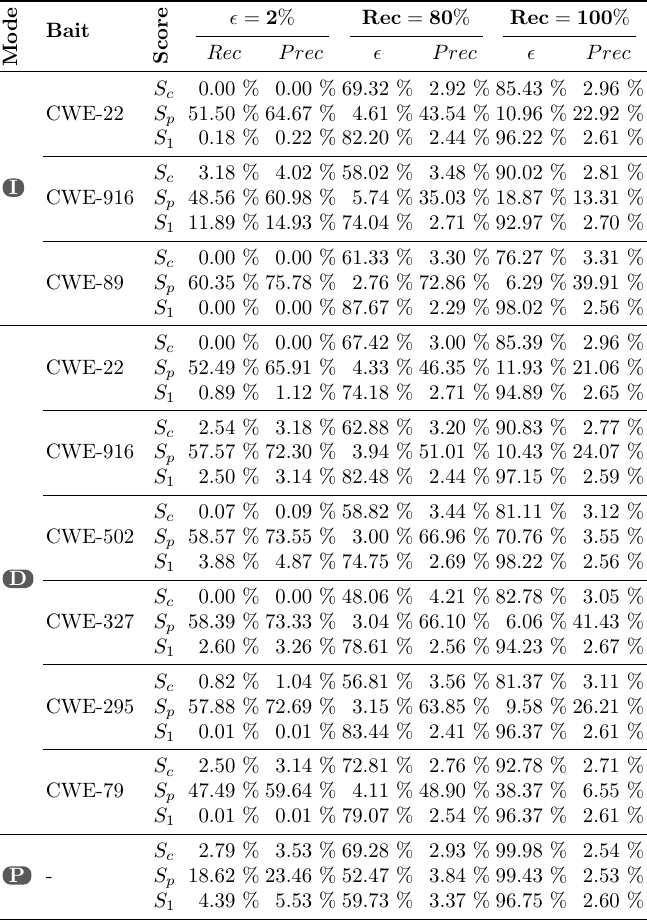}
  }
\end{table}

\end{document}